\documentclass[aps,
prd,
showpacs,nofootinbib,showkeys,10pt]{revtex4-2}
\usepackage[table]{xcolor}
\usepackage{graphicx} \usepackage{amsmath} \usepackage{amssymb}
\usepackage{amsfonts} \usepackage{bm}
\usepackage{array}
\usepackage{subcaption}
\usepackage{siunitx}
\usepackage[singlelinecheck=false]{caption}
\usepackage{ytableau}
\usepackage{multirow}
\usepackage{mathtools}
\usepackage{tikz}
	
\begin{document}
\newcommand{\be}{\begin{equation}} \newcommand{\ee}{\end{equation}}
\newcommand{\bea}{\begin{eqnarray}}\newcommand{\eea}{\end{eqnarray}}

\title{Quantum walk  search  for exceptional configurations on one- and two-dimensional lattice with extra   long-range edges of Hanoi
network of degree four }

\author{Satoshi Watanabe} \email{xsas-watanabe@kddi.com}
\author{Pulak Ranjan Giri} \email{pu-giri@kddi-research.jp}

\affiliation{KDDI Research,  Inc.,  Fujimino-shi, Saitama, Japan}

\begin{abstract} 
There  exist  several  types of configurations of marked vertices, referred to as the exceptional configurations,  on   one- and two-dimensional periodic  lattices with additional long-range edges of the Hanoi
network of degree four (HN4), which are challenging to find  using discrete-time quantum walk algorithms. In this article, we conduct a comparative analysis of the discrete-time  quantum walk algorithm utilizing various coin operators to search for these   exceptional configurations.  First, we study the emergence  of  several  new exceptional configurations/vertices  due to the  additional long-range  edges of the HN4  on both one- and two-dimensional lattices. Second,  our  study shows that  the diagonal configuration on a two-dimensional lattice, which is exceptional in the case without long-range edges, no longer remains an exceptional configuration.  Third, it is also shown that a  recently  proposed modified coin can search all these  configurations, including any other configurations in one- and two-dimensional lattices with very high success probability.  Additionally, we construct stationary states for the exceptional configurations caused by the additional long-range edges,  which explains why the standard and lackadaisical quantum walks with the Grover coin cannot search these configurations.   

\end{abstract}


\date{\today}

\maketitle 



\section{Introduction} \label{in}
A database search \cite{ni} in both classical and quantum contexts is a widely used, crucial task in various applications of computer science. In an unsorted database of size $N$, a classical algorithm  requires $\mathcal{O}(N)$ time  to locate a specific element, called marked or target element,   in the worst-case scenario. In contrast, a quantum algorithm  can perform this task more efficiently, utilizing quantum superposition to its advantage. Specifically, Grover's quantum search algorithm \cite{grover1} can find a marked element in the database in $\mathcal{O}(\sqrt{N})$ time, which represents a quadratic speedup over classical methods. For a comprehensive overview of Grover's search algorithm and its generalizations, please refer to \cite{giri}.
Extending Grover's search to quantum  spatial search, such as search on  graphs \cite{portugal} presents certain challenges. In graph searches, movement is restricted to nearest neighbor vertices in  a single time step.  Simply applying Grover algorithm to  graph search implies that  it will require $\mathcal{O}(\sqrt{N})$ iterations, and  each iteration necessitates an additional $\mathcal{O}(\sqrt{N})$ time for reflection, resulting in a total  $\mathcal{O}(N)$ \cite{beni} time complexity, that is as slow as the classical algorithm.

In this context, the quantum walk---the quantum analog  of the classical random walk---have proven to be effective  in expediting searches on graphs compared to classical exhaustive search method. Two types  of quantum walk methods---continuous-time \cite{childs} and discrete-time \cite{amba2}  based  quantum walk---can be utilized to identify the marked vertices on various graph structures. Notable examples of quantum walk search  include investigations  on one-dimensional \cite{lovet}, two-dimensional \cite{amba2,giri2,childs1}, and higher-dimensional  \cite{childs1} grids  with periodic boundary conditions, among other graphs  \cite{amba4,meyer}. Specifically, in the two-dimensional grid scenario, Grover's algorithm combined with multi-level recursion \cite{amba1} can locate a marked vertex in $\mathcal{O}(\sqrt{N}\log^2 N)$ time, while the discrete-time quantum walk achieves the same task in $\mathcal{O}(\sqrt{N}\log N)$ time, potentially improving to $\mathcal{O}(\log N)$  and further  improvement of $\mathcal{O}(\sqrt{\log N})$ possible with additional techniques \cite{tulsi}. Furthermore, the lackadaisical quantum walk \cite{wong1,wong2,wong3,HøyerYu} can directly search a two-dimensional grid in $\mathcal{O}(\sqrt{N\log N})$ time without requiring supplementary techniques.  In continuous-time quantum walk, an optimal speed of $\mathcal{O} (\sqrt{N})$ for single vertex searches has been achieved \cite{tomo} by incorporating long-range edges into the two-dimensional grid.
In multi-vertex search, optimal performance of    $\mathcal{O} (\sqrt{N/M})$  \cite{giri3} can be reached with the lackadaisical quantum walk when additional long-range edges of  Hanoi
network of degree four (HN4) are added to the two-dimensional grid \cite{giriijqi}. Apart from spatial search, quantum walks can also be applied to other tasks such as edge detection \cite{giripla} of images in image processing, topological phase \cite{Kita} and time-series analysis \cite{Konno}. 

In discrete-time quantum walk search, marked vertices are differentiated from unmarked ones by modifying the coin operator. Two prevalent modified coins,  SKW coin, $\mathcal{C}_{SKW}$, \cite{she} and Grover coin, $\mathcal{C}_{Grov}$, \cite{amba2}, facilitate this differentiation. For instance, the algorithm based on the SKW coin applies the Grover diffusion operator, $D_0$, to the unmarked vertices while applying $-\mathbb{I}$ to the marked vertices. Conversely, the Grover coin-based quantum walk algorithm applies $D_0$ to the unmarked vertices and $-D_0$ to the marked vertices. Search algorithms  utilizing these coins have shown considerable success in identifying a single marked vertex in a graph. However, challenges arise in searching for multiple marked vertices \cite{rivosh,nahi1,saha}. For instance, the Grover coin fails to locate two adjacent marked vertices \cite{nahi2} on a two-dimensional periodic lattice. Additionally, configurations where marked vertices are organized in a $2k \times m$ or $k \times 2m$ block, for any positive $k$ and $m$, cannot be efficiently searched. However, when marked vertices are arranged in a $k \times m$ block with both dimensions as odd numbers, the Grover coin can successfully identify any of the marked vertices within those blocks. Conversely, the SKW coin struggles to find  marked vertices aligned diagonally on a two-dimensional grid and certain generalized configurations \cite{amba5,men}. These configurations, which cannot be searched effectively using the SKW and Grover coins, are collectively referred to as the {\it exceptional configurations}.

To address the limitations associated with the SKW and Grover coins, a recently  proposed    modified coin operator, $\mathcal{C}_{G}$, as detailed in ref. \cite{giri4} has proven to be very useful.  This operator modifies the coin operator $\mathcal{C}_{l}$ from the lackadaisical quantum walk. Instead of flipping the sign of all basis states corresponding to marked vertices, it solely flips the sign of the self-loop attached to the marked vertices, followed by the application of the Grover diffusion operator on the marked vertices' coin basis states. Consequently, the quantum walk search effectively reduces to a Grover search for the self-loops associated with the marked vertices. This newly proposed coin operates successfully with {\it exceptional configurations} as well as any other arrangement of marked vertices.
Research indicates that the SKW and Grover coins, along with the lackadaisical quantum walk coin, $\mathcal{C}_l$, face difficulties in discovering the {\it exceptional configurations} of marked vertices. However, increasing the number of marked vertices tends to make the classical search problem easier, yielding reduced search times and higher success probabilities. This observation  had led to the assertion  \cite{nahi2,wong4} that the emergence of {\it exceptional configurations} belongs to the realm of quantum phenomena.

In this article, we study one- and two-dimensional lattices with extra long-range edges given by the HN4 \cite{boe1,boe2,marq}.  A two-dimensional  periodic lattice with this extra long-range edges is an interesting graph, because it leads to an optimal  time complexity  of  $\mathcal{O} (\sqrt{N/M})$  \cite{giri3} for the quantum search  by the lackadaisical quantum walk. 
We conduct a comparative analysis of quantum walk algorithms utilizing different coin operators in searching for {\it exceptional configurations} on one-dimensional lattices with extra edges and {\it non-exceptional configurations} and {\it exceptional configurations} on two-dimensional lattices with extra edges. In one- and two-dimensional lattices, our findings are that the configuration of two marked points adjacent by long-range edges and the configuration of  points which have directed  self-loop  of long-range edges are new exceptional configurations. We numerically demonstrate  that these configurations cannot be searched by $\mathcal{C}_{Grov}$ and $\mathcal{C}_{l}$, and we construct stationary states associated with these configuration which explain why $\mathcal{C}_{Grov}$ and $\mathcal{C}_{l}$ cannot search these configurations. 
In the two-dimensional case, one of our finding is that diagonal configurations which were  exceptional  without  long-range edges(regular two-dimensional periodic lattice) are not exceptional configuration when long-range edges are attached to the lattice.  With  long-range edges, movement along long-range  edges are allowed, and  coin  states  associated with the long-range  edges exist, which do not allow to form the  stationary states associated with diagonal configurations.  Therefore,  diagonal configurations can be  searched by four coins in two-dimensional lattice with extra long-range edges. 
Furthermore, our  experiment  indicates  that the $\mathcal{C}_{G}$ coin can search these new exceptional configurations along with other exceptional configurations   with high success probabilities. 

The organization of this article is as follows: Section \ref{1D} provides a brief overview of quantum walk search  on a one-dimensional periodic lattice with extra long-range edges along with the  numerical  analysis   of searching  for {\it exceptional configurations} using  various coin operators.
Section \ref{2D} provides a brief overview of quantum walk search  on a two-dimensional periodic lattice with extra long-range  edges along with the  numerical  analysis of  searching  for {\it non-exceptional configurations} and {\it exceptional configurations}.  And finally   we conclude in Section \ref{con}.

\section{One-dimensional periodic lattice with extra long-range edges} \label{1D}

\begin{figure}[h!]
    \centering
    \includegraphics[scale=0.25]{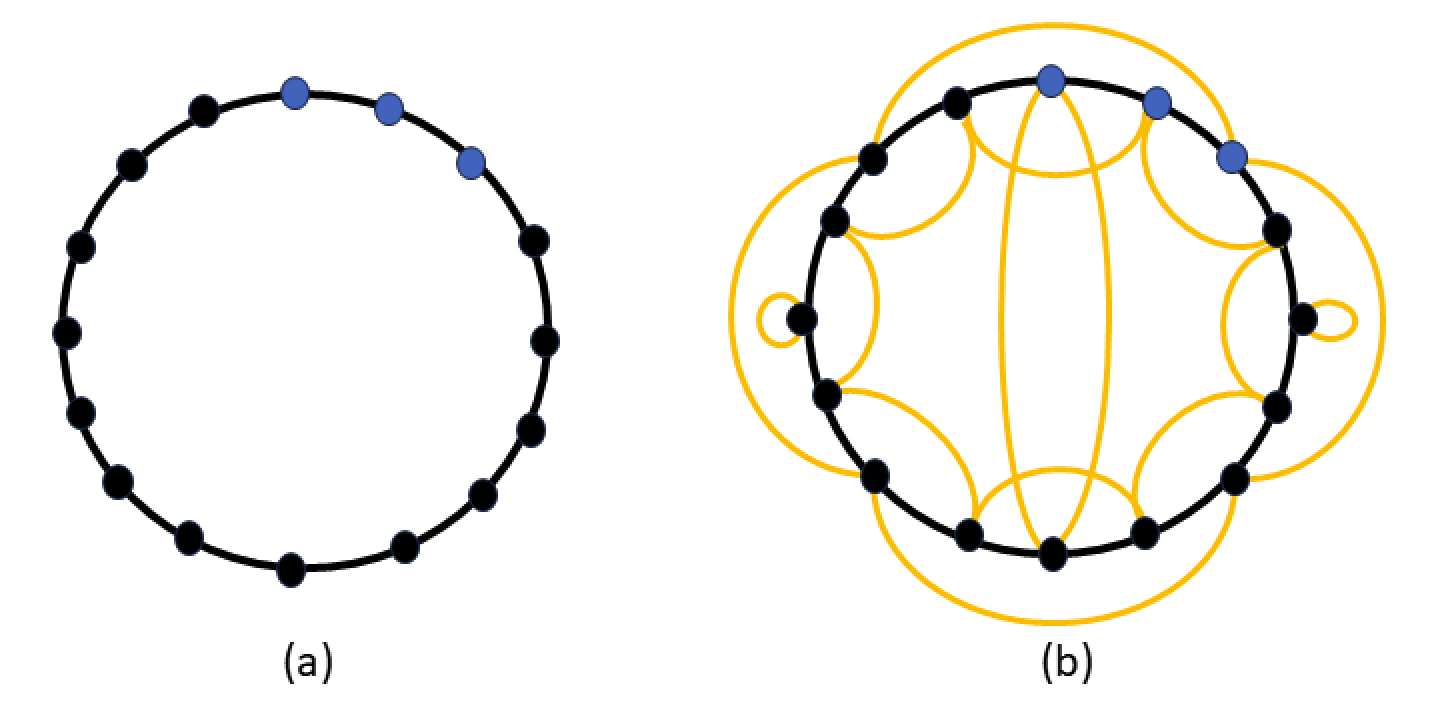}
 \caption{ (a) One-dimensional periodic  lattice of size $N=16$  with three  blue colored  marked vertices and (b) Hanoi network of degree four (HN4)  with  $N=16$  vertices and yellow colored long-range edges.}
 \label{fig1}
\end{figure}

In this section, we present a concise overview of the discrete-time quantum walk search algorithm on a one-dimensional periodic  lattice of size $N$  that has additional  long-range edges  given by the Hanoi
network of degree four (HN4) \cite{boe1}. In FIG. \ref{fig1}(a) a one-dimensional periodic lattice with $N=16$   vertices,  has three  marked vertices  represented with  blue color.  And its  corresponding HN4 is presented in  FIG.   \ref{fig1}(b).  Note that HN4 has yellow colored long-range  edges apart from the black colored standard/regular  edges.  So, in HN4 ,   each vertex is associated with two standard/regular  edges represented by the basis states $|0 \rangle$ and $| 1 \rangle$, as well as two long-range  edges described by the basis states $| 2 \rangle$ and $|3  \rangle$. 
The Hilbert space of the graph, denoted as $\mathcal{H}_{G}=\mathcal{H}_{C}\otimes \mathcal{H}_{V}$, is the tensor product of the coin space $\mathcal{H}_C$ and the vertex space $\mathcal{H}_V$. 
In terms of the coordinates $(i, j)$  of the HN4 , each vertex  $1 \le v_{x} \le N= 2^{n}$ of the one-dimensional lattice  is given by the following equation:
\begin{align}
v_{x}=2^{i}(2 j+1)\,,~~  \mbox{for}~~~~ 0 \le i \le n; ~~~ 0 \le j \le j_{\mathrm{max}} = \lfloor 2^{n-i-1} - 1/2 \rfloor\,.
\end{align}
The basis state  $|v_{x}\rangle$   for the lattice can  alternatively  be expressed  in terms of the HN4 coordinates  as  $|i,j\rangle$.  Note that, vertices  $(i, j) =(n-1, 0)$ and $(n, 0)$ corresponding to Cartesian coordinates   $N/2$ and $N$  on the one-dimensional lattice form  directed self-loops, such as the  two yellow colored self-loops on left and right side  of   FIG. \ref{fig1}(b). These two locations  are  exceptional points, as we will see  these vertices are hard to find.  

The quantum walk search begins with an initial state of the form
\begin{eqnarray}
|\psi_{1d}(0)\rangle = |\psi_{c}(0) \rangle \otimes |\psi_{v}(0) \rangle\,,
\label{in1d}
\end{eqnarray}
where both the coin and vertex spaces are initialized in a uniform superposition of all the  basis states as
\begin{align}
&|\psi_{c}(0) \rangle =\frac{1}{\sqrt{4}} \sum^{3}_{{v_c} = 0 }|v_c\rangle\,,\\
&|\psi_{v}(0)\rangle =   \frac{1}{\sqrt{N}} \sum^{N}_{{v_x} = 1} |v_x\rangle .
\end{align}
Note that eq. (\ref{in1d}) is the initial state for the quantum walk search algorithm without self-loop, such as for Grover and SKW coins.  The time evolution operator $\mathcal{U}$ acts on the initial state $|\psi_{1d}(0) \rangle$ repeatedly, leading to a final state
\begin{align}
|\psi_{1d}(t) \rangle =\mathcal{U}^{t}|\psi_{1d}(0)\rangle\,,
\end{align}
that  aims to achieve high fidelity with the state representing the marked vertices. In the case of lackadaisical quantum walk, i.e., quantum walk with self-loop,  the final state $|\psi_{1d}(t)\rangle$ directly corresponds to the state of the marked vertices with high success  probability. However, in the quantum walk search without self-loops, the final state generally does   not exhibit significant overlap with the state of  marked vertices, necessitating the use of amplitude amplification techniques \cite{brassard} to ensure a high and constant  success probability.

The time evolution operator in the quantum walk search algorithm is defined as the composition of the coin operator $\mathcal{C}$ and the flip-flop shift operator $S$:
\begin{align}
 \mathcal{U}=S \mathcal{C}\,.
\end{align}
The  generalized coin operator  $\mathcal{C} = \Lambda_{un} +  \Lambda_{m}$ has two parts.   $\Lambda_{un}$ acts    on  the coin space associated with the  unmarked  vertices  and   $\Lambda_{m}$ acts on the coin space 
associated with the marked vertices  respectively. $M$ is defined as the number of the marked vertices. Formally  they are   given by  
\begin{eqnarray}\nonumber 
\Lambda_{un} &=&  \Lambda_+ \otimes \left( \mathbb{I}_{N \times N} -    \sum_{i=1}^M |t_i \rangle \langle t_i |  \right)\,,\\
\Lambda_{m}  &=&   \Lambda_- \otimes  \sum_{i=1}^M |t_i \rangle \langle t_i | \,, 
\label{qc1d}
\end{eqnarray}
where $\Lambda_+$ and $\Lambda_-$ are the coin operators acting on  unmarked and marked vertices  respectively, and $|t_i \rangle$s  denote  the marked vertex states. 
Grover coin $\mathcal{C}_{Grov}$  can be  obtained  from the  expression in eq. (\ref{qc1d}), by replacing 
\begin{eqnarray}\nonumber 
\Lambda_+ &=&  D_0\,,\\
\Lambda_- &=&  - D_0\,,
\label{qc1dg}
\end{eqnarray}
where  $D_0 = 2|\psi_{c}(0) \rangle \langle \psi_{c}(0)| - \mathbb{I}_{4 \times 4}$ is the Grover diffusion operator. Similarly,  SKW coin  $\mathcal{C}_{SKW}$  can  be obtained   from eq. (\ref{qc1d}), by replacing 
\begin{eqnarray}\nonumber 
\Lambda_+ &=&  D_0\,,\\
\Lambda_- &=&  -  \mathbb{I}_{4 \times 4}\,.
\label{qc1dskw}
\end{eqnarray}
$\mathcal{C}_{SKW}$ was  first utilized by Shenvi et al. \cite{she} to search for a marked vertex on a hypercube and a complete graph. It's worth noting that above mentioned  two coins are  referred to as Grover and AKR coins in the context of  AKR algorithm \cite{nahi2,giri4,amba2}.

The quantum walk can be extended to a lackadaisical quantum walk by incorporating an  undirected self-loop state $|\mbox{lp} \rangle$ at each vertex of the graph, weighted by $l$. 
The initial state in this case is given by 
\begin{eqnarray}
|\psi_{1d}(0)\rangle = |\psi_{cl}(0) \rangle \otimes |\psi_{v}(0) \rangle\,,
\label{in1dl}
\end{eqnarray}
where the  initial state for the coin space is expressed as
\begin{align}
&|\psi_{cl}(0) \rangle =\frac{1}{\sqrt{4+l}}\Bigl( \sum^{3}_{{v_c} = 0 }|v_c\rangle +\sqrt{l} |\mbox{lp} \rangle \Bigr).
\label{in1dlc}
\end{align}
Note that eq. (\ref{in1dl}) is the initial state for the quantum walk search algorithm with  self-loop, such as for lackadaisical quantum walk with $\mathcal{C}_l$ and $\mathcal{C}_G$ coins.  The coin operator for the lackadaisical quantum walk search,  $\mathcal{C}_l$,    can be  obtained  from the  expression in eq. (\ref{qc1d}), by replacing 
\begin{eqnarray}\nonumber 
\Lambda_+ &=&  D_l\,,\\
\Lambda_- &=&  - D_l\,,
\label{qc1dgl}
\end{eqnarray}
where the Grover diffusion operator for the lackadaisical quantum walk  $D_{l} = 2|\psi_{cl}(0) \rangle \langle \psi_{cl}(0)| - \mathbb{I}_{5 \times 5}$ is derived from the initial state in eq. (\ref{in1dlc}). 
It is important to note that the quantum walk with coin $\mathcal{C}_{Grov}$ can be retrieved from the lackadaisical variant by setting $l = 0$. Typically, it is necessary to determine an optimal value for the self-loop weight $l$ to maximize the success probability. However, in this paper, we will simply select a suitable value for $l$ that ensures a reasonably high success probability for our analysis of exceptional configurations.

The limitations of the previously mentioned coin operators in the search for exceptional configurations of marked vertices can be addressed by  a  recently proposed  modified coin operator $\mathcal{C}_{G} = \mathcal{D}_{un;l} + \mathcal{D}_{m;l}$,  where  the operator acting  on unmarked and marked  vertex with attached self-loop  states are   respectively  given by 
\begin{eqnarray}\nonumber 
\mathcal{D}_{un;l} &=&  \left(D_l \otimes \mathbb{I}_{N \times N} \right)  \left( \mathbb{I}_{5 \times 5}  \otimes \mathbb{I}_{N \times N}-    | \mbox{lp} \rangle \langle  \mbox{lp}  |  \otimes \sum_{i=1}^M  | t_i \rangle \langle   t_i |  \right)\,, \\
\mathcal{D}_{m;l} &=&  - \left(D_l \otimes \mathbb{I}_{N \times N} \right)  | \mbox{lp} \rangle \langle  \mbox{lp}  |  \otimes \sum_{i=1}^M  | t_i \rangle \langle   t_i | \,.
\label{qc4}
\end{eqnarray}

The flip-flop shift operator $S=S_{se}+S_{le}+S_{lp}$ is composed of three components, each corresponding to different types of edges. 
\begin{enumerate}
\item $S_{se}$:  The portion related to the two standard  edges of the one-dimensional lattice is expressed as
\begin{align} 
S_{se}=&\sum_{v_{x}=1}^{N} \biggl[ |1 \rangle \langle 0| \otimes |v_{x}+1\rangle \langle v_{x}|+
|0 \rangle \langle 1 |\otimes |v_{x}-1\rangle \langle v_{x}| \biggr],
\label{fshift1}
\end{align}
where the coin basis states $|0 \rangle$ and $|1 \rangle$ correspond to the right and left edges  respectively. 

\item $S_{le}$: The shift operator associated with the two long-range edges at each vertex is represented by
\begin{align} 
S_{le}=&\sum_{i=0}^{n-2}\sum_{j=0}^{j_{max}}\biggl( |3 \rangle \langle 2 |\otimes |i,j+1\rangle \langle i,j|+
|2 \rangle \langle 3 |\otimes |i,j-1\rangle \langle i,j|\biggr) \notag \\
+&\biggl( |3 \rangle \langle 2 |+
|2 \rangle \langle 3 |\biggr)\otimes \biggr(| N/2\rangle \langle N/2|+| N\rangle \langle N |\biggr)\,,
\label{fshift1}
\end{align}
where the coin basis states $|2 \rangle$ and $|3 \rangle$ are associated with the right and left long-range edges  respectively. 

\item $S_{lp}$: Finally, the shift operator corresponding to the undirected self-loop in the lackadaisical quantum walk is given by
\begin{align}
S_{lp}=&\sum_{v_{x}=1}^{N} |\mbox{lp}\rangle \langle \mbox{lp}| \otimes |v_{x}\rangle \langle v_{x}|.
\end{align}
\end{enumerate}
The action of  the flip-flop shift operator  on the basis states of the tensor product of coin  and vertex space can be summarized  as  
\begin{eqnarray}
S| 0 \rangle \otimes | v_x \rangle &=& S_{se}| 0 \rangle \otimes | v_x \rangle  = | 1 \rangle \otimes | v_x + 1\rangle\,,  \\
S| 1 \rangle \otimes | v_x \rangle  &=&  S_{se}| 1 \rangle \otimes | v_x \rangle  = | 0 \rangle \otimes | v_x - 1\rangle\,, \\
S| 2 \rangle \otimes | i,j \rangle &=& S_{le}| 2 \rangle \otimes | i,j \rangle = | 3 \rangle \otimes | i,j+1\rangle\,, \\
S| 3 \rangle \otimes | i,j \rangle  &=&  S_{le}| 3 \rangle \otimes | i,j \rangle = | 2 \rangle \otimes | i,j-1\rangle\,, \\
S| \mbox{lp} \rangle \otimes | v_x\rangle &=&  S_{lp}| \mbox{lp} \rangle \otimes | v_x\rangle = | \mbox{lp} \rangle \otimes | v_x \rangle\,.
\label{2Dcgrov}
\end{eqnarray}
The success probability to find the marked vertices becomes
\begin{align}
   p_{1d}(t)=\sum_{i=1}^{M}\Bigl(\sum^{3}_{{v_c} = 0 }|\langle v_c | \otimes \langle t_{i}|  \mathcal{U}^{t}|\psi_{1d}(0) \rangle |^{2} +|\langle  \mbox{lp} | \otimes \langle t_{i}|  \mathcal{U}^{t}|\psi_{1d}(0) \rangle |^{2}\Bigr),
\end{align}
where sum over coin basis states is assumed to be  done at the marked vertices.

\subsection{Exceptional configurations for the one-dimensional periodic lattice with extra long-range edges} \label{1Ds}

Usually, in the one-dimensional periodic lattice without any long-range edges  there is no exceptional configuration. However, the  additional  long-range  edges of the HN4 creates some  exceptional configurations.  As previously mentioned, these  exceptional configurations  occur  due to  the directed self-loops at  $(n-1, 0) = N/2$ and  $(n,0)= N$. In this subsection, we  discuss these  configurations in detail and show how   searching  these  configurations  by $\mathcal{C}_{SKW}$  and $\mathcal{C}_{G}$ coins fails. We also discuss the existence of stationary states that have negligible  overlap with the initial state of uniform superposition of all the basis states. These stationary states  are  responsible for the failure to find these  exceptional configurations  by the $\mathcal{C}_{Grov}$  and  $\mathcal{C}_{l}$  coin based quantum walk search algorithms.
\begin{figure}[h!]
    \centering
    \includegraphics[scale=0.60]{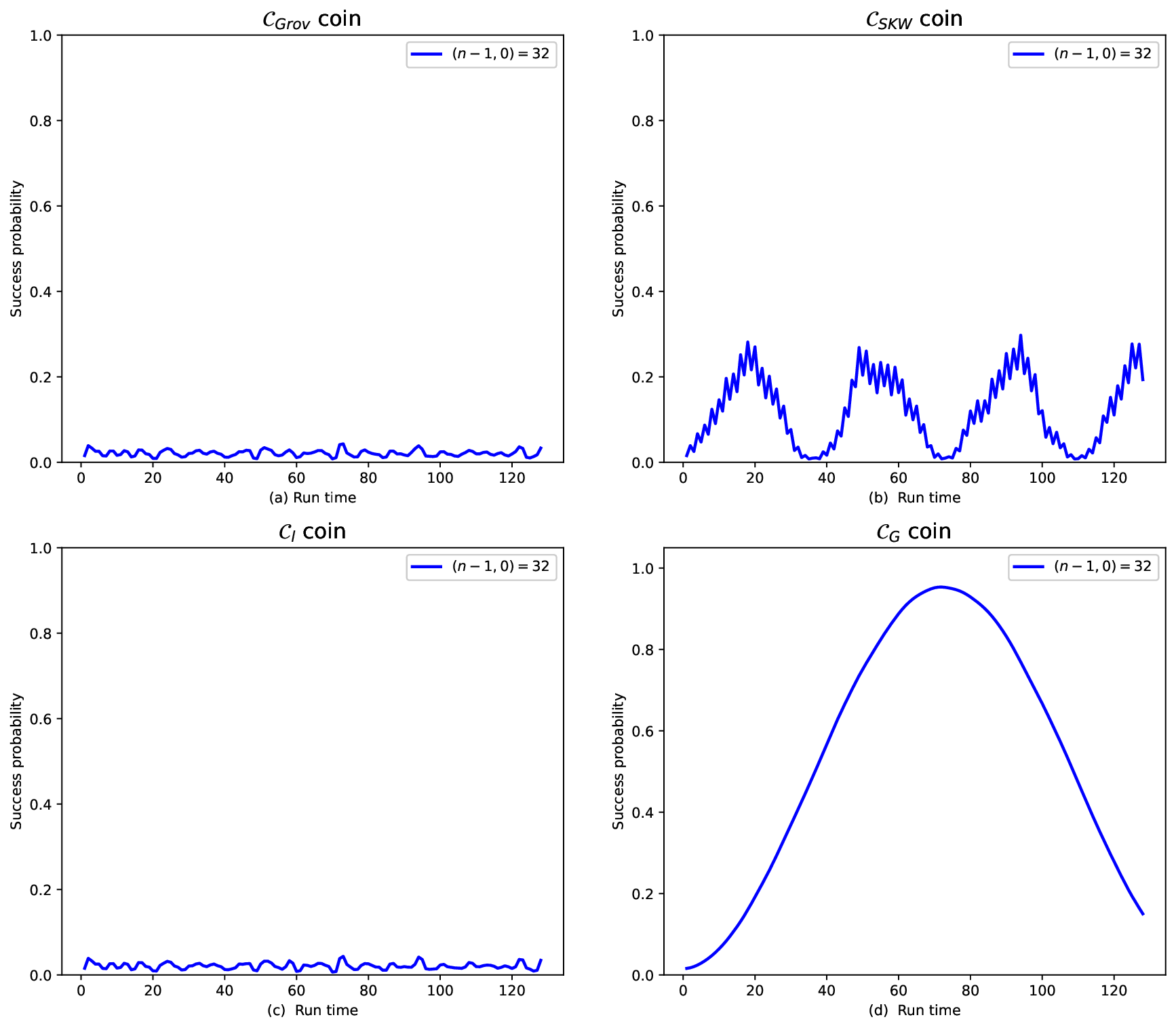}
  \caption{Success probability to measure the  marked vertex $(n-1,0)=(5,0)= 32$ in $N=64$ one-dimensional periodic lattice with extra long-range  edges as a function of the number of iteration steps obtained using  (a) $\mathcal{C}_{Grov}$, (b) $\mathcal{C}_{SKW}$, (c) $\mathcal{C}_{l}$, and (d) $\mathcal{C}_{G}$ coins.}
  \label{1D-exceptional}
\end{figure}

\subsubsection{Searching  for vertices  with directed self-loops of the long-range edges}

{\it Experimental results:} The behavior of the success probability to measure a marked vertex that  has a directed self-loop at $(n-1,0) = (5, 0) = 32$ on a $N=64$ one-dimensional lattice as a function of the number of iteration steps has been presented in FIG. \ref{1D-exceptional}.  Self-loop weight associated with the lackadaisical quantum walk is fixed at $l= 2.0/N$.   
We see that the success probability for  $\mathcal{C}_{Grov}$ and $\mathcal{C}_{l}$ coins represented by (a) and (c) do not grow at all even if we increase the number of iteration steps. For the $\mathcal{C}_{SKW}$ and $\mathcal{C}_{G}$ coins,  success probability grow as a function of the number of iteration steps represented by (b) and (d). For $\mathcal{C}_{SKW}$ coin, as we can see, the success probability is not significantly high, so we need to further apply amplitude amplification \cite{brassard} to enhance the success probability. The experiment is also carried out for the vertex at  $(n,0) =(6,0) = 64$ that  gives us the   same conclusion.  The reason that these configurations cannot be searched by $\mathcal{C}_{Grov}$ and $\mathcal{C}_{l}$ coins is explained theoretically by the formation of  stationary states.

\subsubsection{Stationary state}
We examine the stationary states, that  exhibit  substantial overlap with the initial state of the quantum walk search. The methodology for constructing the stationary state corresponding to adjacent marked vertices with respect to standard edges, as well as its application to the quantum walk search, is detailed in refs. \cite{nahi1,nahi2,wong4}. In this sub-subsection, we focus on constructing the stationary state for the exceptional configuration at a marked  vertex that is associated with  a  directed self-loop formed by a long-range edge. 
The stationary state is defined as the eigenvector of the coin operator with a unit eigenvalue. This vector is crucial for understanding why  $\mathcal{C}_{Grov}$ and  $\mathcal{C}_{l}$ are ineffective for searching  these  exceptional configurations. To describe the action of the coin operators on the stationary state, we introduce the following pair of non-normalized coin states
\begin{eqnarray}
|\psi_{c}^{\mathrm{le},+} \rangle =|\psi_{cl}(0) \rangle  - \sqrt{4+l}|2 \rangle,
~~~|\psi_{c}^{\mathrm{le},-} \rangle =|\psi_{cl}(0) \rangle  - \sqrt{4+l}|3 \rangle\,.
\end{eqnarray}
It is important to note that these two states are orthogonal to $|\psi_{cl}(0) \rangle$, i.e., $\langle \psi_{c}^{\mathrm{le},+}|\psi_{cl}(0) \rangle = \langle \psi_{c}^{\mathrm{le},-}|\psi_{cl}(0) \rangle = 0$.
We consider the stationary state associated with the exceptional vertex  $(n-1,0) = 2^{n-1} =N/2$ as
\begin{align}
|\psi_{HNstat} \rangle =|\psi_{1d}(0) \rangle - \sqrt{\frac{4+l}{N}} \frac{1}{2}\biggl( |2  \rangle + |3  \rangle \biggr) \otimes  |N/2 \rangle,
\end{align}
where the initial state  $|\psi_{1d}(0) \rangle$  is given by eq. (\ref{in1dl}).
Note that, the stationary state for the exceptional point $(n,0) = 2^{n}= N$ can be  defined in a similar manner. The action of the four coin operators on this stationary  state is summarized as follows:  
\begin{enumerate}
\item $U_{l}$: It can be shown  that the stationary state $|\psi_{HNstat} \rangle$ is an eigenstate of the time evolution operator, $U_{l} = S\mathcal{C}_{l}$,  associated with the lackadaisical quantum walk  with a unit eigenvalue:
\begin{align}
    U_{l} |\psi_{HNstat} \rangle =|\psi_{HNstat} \rangle.
\end{align}

\item $U_{Grov}$:  $|\psi_{HNstat} \rangle$  remains an eigenstate for the regular quantum walk search with $U_{Grov} = U_{l=0} = S\mathcal{C}_{Grov}$ coin,
\begin{align}
    U_{Grov} |\psi_{HNstat} \rangle =|\psi_{HNstat} \rangle.
\end{align}

\item $U_{G}$: However, under  $U_{G} = S \mathcal{C}_{G}$, the stationary state does not remain stationary, as illustrated by the following expression
\begin{align}
    U_{G} |\psi_{HNstat} \rangle =|\psi_{HNstat} \rangle-\frac{1}{\sqrt{N}}S\biggl[\Bigl(|\psi_{c}^{\mathrm{le},+} \rangle +|\psi_{c}^{\mathrm{le},-} \rangle \Bigr) \otimes  |N/2 \rangle \biggr] -2\sqrt{\frac{l}{N(4+l)}}S D_{l}\biggl(|\mbox{lp} \rangle \otimes |N/2 \rangle \biggr).
\end{align}

\item $U_{SKW}$: Similarly, under $U_{SKW} = S \mathcal{C}_{SKW}$, the stationary state also does not remain stationary, as shown by the following expression:
\begin{align}
    U_{SKW} |\psi_{HNstat} \rangle =|\psi_{HNstat} \rangle-\frac{1}{\sqrt{N}}\mathcal{S}\biggl[\Bigl(|\psi_{c}^{\mathrm{le},+} \rangle +|\psi_{c}^{\mathrm{le},-} \rangle \Bigr) \otimes  |N/2 \rangle  \biggr]
\end{align}
Note that, for the quantum walk search with $\mathcal{C}_{Grov}$, $\mathcal{C}_{SKW}$ coins there is no self-loop, so we need to set $l =0$ wherever necessary. 
\end{enumerate}
Let us now  express the initial state $|\psi_{1d}(0) \rangle$ in terms of the stationary state $|\psi_{HNstat} \rangle$ as 
\begin{align}
|\psi_{1d}(0) \rangle =|\psi_{HNstat} \rangle + \sqrt{\frac{4+l}{N}} \frac{1}{2}\biggl( |2 \rangle + |3  \rangle \biggr) \otimes |N/2 \rangle.
\label{initialof1d}
\end{align}
The final state, after applying $U_{l}$ repeatedly $t$ times,  becomes 
\begin{align}
U_{l}^{t}|\psi_{1d}(0) \rangle  =|\psi_{HNstat} \rangle + \sqrt{\frac{4+1}{N}} \frac{1}{2}U_{l}^{t}\biggl[\Bigl( |2 \rangle + |3  \rangle \Bigr) \otimes |N/2 \rangle\biggr].
\label{time1d}
\end{align}
It is noteworthy that the first component on the right side of eq.  \eqref{time1d} remains unchanged under the action of the evolution operator, while  the second component is affected. The upper bound for the success probability of locating the  marked vertex can be determined by setting $U_{l}^{t} = -\mathbb{I}$ in the second term. Thus, the success probability for the  marked vertex is constrained by  $p_{1d}(t) \le \mathcal{O}(1/N)$. The same argument is also applicable to $U_{Grov}$. It indicates that the quantum walk search using $U_{l}$ and $U_{Grov}$ coins on a one-dimensional periodic lattice with extra long-range  edges cannot locate the vertex with directed self-loop, as the success probability remains limited by the initial success probability. 
However, under the action of $U_{G}$ on the initial state, both the first and second components of eq.  \eqref{initialof1d} transform non-trivially, since $|\psi_{HNstat}\rangle$ is no longer an eigenstate of $U_{G}$. After the repeated application of the evolution operator, the final state achieves a significantly high and constant overlap with the marked vertex.
In the case of $U_{SKW}$, both components of the initial state also transform non-trivially. However, as mentioned before, to achieve a high success probability, we must implement amplitude amplification in addition to the repeated application of $U_{SKW}$.

\section{Two-dimensional periodic lattice with extra long-range  edges} \label{2D}
\begin{figure}[h!]
    \centering
    \includegraphics[scale=0.25]{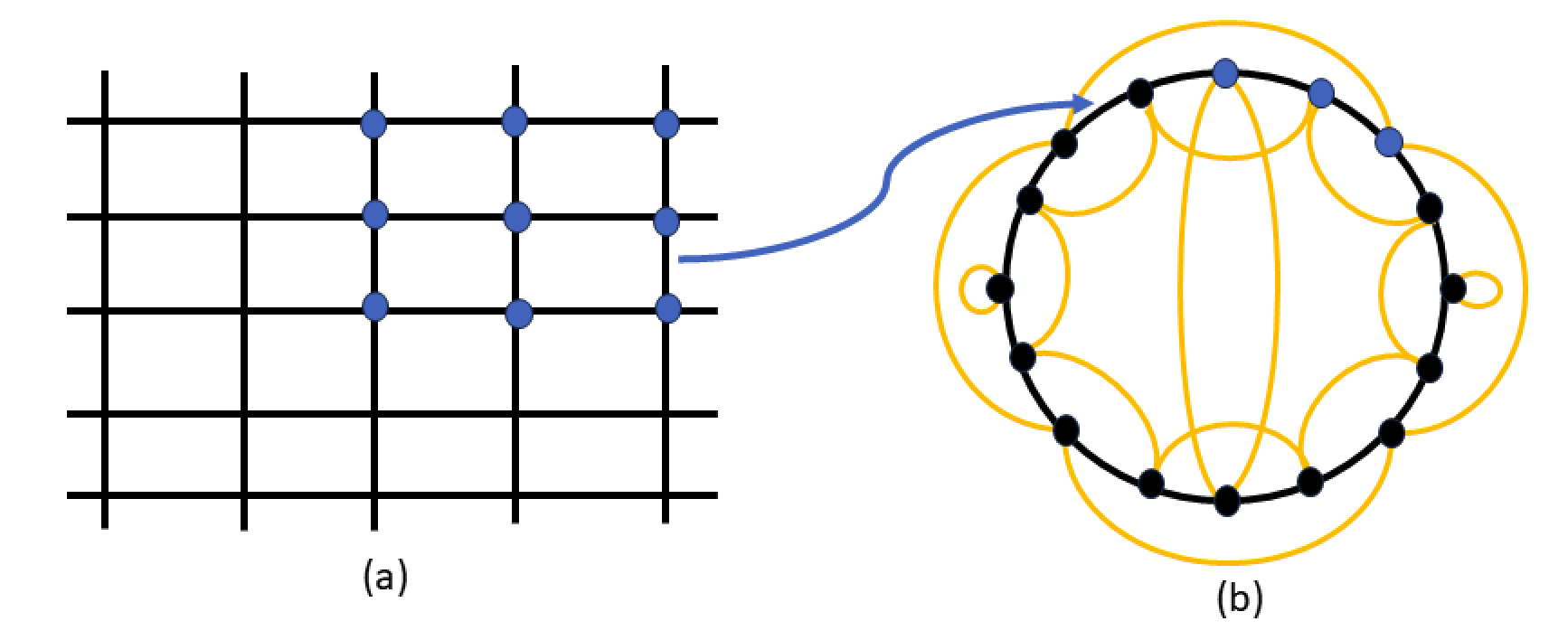}
    \label{2D-exceptional}
 \caption{ (a) Two-dimensional lattice of size $\sqrt{N} \times \sqrt{N}$ with periodic boundary conditions with a blue  colored cluster of marked vertices and (b) Each row and column of the lattice in left panel is additionally added with long-range edges of Hanoi network of degree four as depicted in right panel for a $16\times 16$ square lattice. In lackadaisical quantum walk search, self-loop is also added at each vertex of the two-dimensional lattice.}
\label{2dlattice}
\end{figure}

In this section, we present a concise overview of the discrete-time quantum walk search algorithm on a two-dimensional lattice of size $\sqrt{N} \times \sqrt{N}$ that includes extra  long-range edges of the HN4  with  periodic boundary conditions on each row and column of  the lattice.  In FIG. \ref{2dlattice}(a) a $5 \times 5$ portion of a $16 \times 16$ periodic $2$-dimensional    lattice is displayed. Each vertex is associated with four standard/regular   edges represented by the basis states $|0 \rangle, |1 \rangle, |2 \rangle$ and $|3  \rangle$, as well as four long-range edges described by the basis states $|4  \rangle, |5 \rangle, |6 \rangle$ and $|7 \rangle$ within the coin space. The Hilbert space of the graph, denoted as $\mathcal{H}_{G}=\mathcal{H}_{C}\otimes \mathcal{H}_{V}$, is the tensor product of the coin space $\mathcal{H}_C$ and the vertex space $\mathcal{H}_V$. 
In terms of the coordinates of the HN4, each vertex  $1\le v_{x},v_{y} \le  \sqrt{N} =2^{n}$  of the two-dimensional lattice  is  defined as follows: 
\begin{align}
    v_{x}=2^{i_{1}}(2 j_{1}+1)\,,~  \mbox{for}~~~0 \le i_{1} \le n,\; 0 \le j_{1} \le j_{1_{\mathrm{max}}}=\lfloor 2^{n-i_{1}-1}-1/2 \rfloor \\
    v_{y}=2^{i_{2}}(2 j_{2}+1)\,,~   \mbox{for}~~~0 \le i_{2} \le n,\; 0 \le j_{2} \le j_{2_{\mathrm{max}}}=\lfloor 2^{n-i_{2}-1}-1/2 \rfloor 
\end{align}
The basis state $ |v_{x};v_{y}\rangle$  can alternatively be   expressed  in terms of the HN4 coordinates as  $|i_{1},j_{1};i_{2},j_{2} \rangle $. 
There are directed self-loops for $v_x= (i_1, j_1) = (n-1, 0)= \sqrt{N}/2$  and  for   $v_x= (i_1, j_1) =(n, 0)= \sqrt{N}$ along along the horizontal directions, and similarly along the vertical directions    i.e.,  at $|\sqrt{N}/2; v_{y}\rangle$, $|\sqrt{N}; v_{y}\rangle$ and $|v_{x}; \sqrt{N}/2\rangle$, $|v_{x}; \sqrt{N} \rangle$ respectively , which are exceptional configurations arise  due to the self-loop formation by  some long-range edges.

The quantum walk search begins with an initial state of the form
\begin{eqnarray}
|\psi_{2d}(0)\rangle = |\psi_{c}(0) \rangle \otimes |\psi_{v}(0) \rangle\,,
\label{in2d}
\end{eqnarray}
where both the coin and vertex spaces are initialized in a uniform 
superposition of all the basis states of the respective spaces as
\begin{eqnarray} \label{in2d1}
|\psi_{c}(0) \rangle &= & \frac{1}{\sqrt{8}}\sum_{{v_c} = 0}^{7} |v_c \rangle\,,  \\ 
|\psi_{v}(0)\rangle  &= &  \frac{1}{\sqrt{N}}  \sum^{N}_{{v_x, v_y} = 1} |v_x;  v_y\rangle\,.
\label{in2d2}
\end{eqnarray}
Note that eq. (\ref{in2d}) is the initial state for the quantum walk search algorithm without self-loop, such for quantum walk   with  $\mathcal{C}_{Grov}$ or $\mathcal{C}_{SKW}$ coins. 
The time evolution operator $\mathcal{U}$ then acts on the initial state $|\psi_{2d}(0) \rangle$ repeatedly until the final state
\begin{align}
|\psi_{2d}(t) \rangle =\mathcal{U}^{t}|\psi_{2d}(0) \rangle ,
\end{align}
has high fidelity with the state representing the marked vertices. 

For the two-dimensional lattice,  Grover coin $\mathcal{C}_{Grov}$  can be  obtained  from the  expression in eq. (\ref{qc1d}), by replacing 
\begin{eqnarray}\nonumber 
\Lambda_+ &=&  D_0\,,\\
\Lambda_- &=&  - D_0\,,
\label{qc1dg}
\end{eqnarray}
where the Grover diffusion operator  $D_0 = 2|\psi_{c}(0) \rangle \langle \psi_{c}(0)| - \mathbb{I}_{8 \times 8}$.  
Here,  initial coin state  $|\psi_{c}(0) \rangle$ is given by  eq. (\ref{in2d1}).
Similarly, $\mathcal{C}_{SKW}$ can  be obtained   from eq. (\ref{qc1d}), by replacing 
\begin{eqnarray}\nonumber 
\Lambda_+ &=&  D_0\,,\\
\Lambda_- &=&  -  \mathbb{I}_{8 \times 8}\,.
\label{qc1dskw}
\end{eqnarray}
In two-dimensions  also to  generalize quantum walk  to lackadaisical quantum walk, we  add  self-loop with weight $l$ to each vertex of the graph. This argument is also parallel to  the case of one-dimensional lattice. 
The initial state is given by 
\begin{eqnarray}
|\psi_{2d}(0)\rangle = |\psi_{cl}(0) \rangle \otimes |\psi_{v}(0) \rangle\,,
\label{in2dl}
\end{eqnarray}
where the  initial state for  the coin space is given by 
\begin{align}
&|\psi_{cl}(0) \rangle =\frac{1}{\sqrt{8+l}}\Bigl(\sum_{{v_c} = 0}^{7} |v_c \rangle+\sqrt{l} | \mbox{lp}\rangle \Bigr).
\label{in2dlc}
\end{align}
Note that eq. (\ref{in2dl}) is the initial state for the quantum walk search algorithm with  self-loop, such as  for  $\mathcal{C}_{l}$ or  $\mathcal{C}_{G}$ coins.  
The coin operator,  $\mathcal{C}_l$, for the lackadaisical quantum walk search,  defined on two-dimensional lattice,    can be  obtained  from the  expression in eq. (\ref{qc1d}), by replacing 
\begin{eqnarray}\nonumber 
\Lambda_+ &=&  D_l\,,\\
\Lambda_- &=&  - D_l\,,
\label{qc2dgl}
\end{eqnarray}
where the Grover diffusion operator $D_{l}=2|\psi_{cl}(0) \rangle \langle \psi_{cl}(0)|- \mathbb{I}_{9 \times 9}$ can be obtained from the initial state in  eq. (\ref{in2dlc}). 
Note that, regular quantum walk can be obtained from lackadaisical quantum walk by setting $l=0$. 
In two-dimensional case,  $\mathcal{C}_{G}$ coin is defined as follows  
\begin{eqnarray}
\mathcal{C}_{G}=  \left(D_l \otimes \mathbb{I}_{N \times N} \right)  \left( \mathbb{I}_{9 \times 9}  \otimes \mathbb{I}_{N \times N}-   2 | \mbox{lp} \rangle \langle  \mbox{lp}  |  \otimes \sum_{i=1}^M  | t_i \rangle \langle   t_i |  \right)\,.
\label{qc4}
\end{eqnarray}
The flip-flop shift operator,  $S=S_{se}+S_{le}+S_{lp}$, is composed of three parts which are associated with three different types of edges. 
\begin{enumerate}
\item $S_{se}$: It is associated with four standard   edges of the $2$-dimensional lattice as 
\begin{align} 
S_{se}=&\sum_{v_{y}=1}^{\sqrt{N}} \sum_{v_{x}=1}^{\sqrt{N}} \biggl[ |1 \rangle \langle 0 | \otimes |v_{x}+1;v_{y}\rangle \langle v_{x};v_{y}|+
|0 \rangle \langle 1 |\otimes |v_{x}-1;v_{y}\rangle \langle v_{x};v_{y}| \notag \\
&\;\;\;\;\;\;\;\;\;\;+|3  \rangle \langle 2 |\otimes |v_{x};v_{y}+1\rangle \langle v_{x};v_{y}|
+|2 \rangle \langle 3 |\otimes |v_{x};v_{y}-1\rangle \langle v_{x};v_{y}|\biggr],
\label{fshift1}
\end{align}
where the coin basis states $|0  \rangle, |1  \rangle, |2 \rangle$ and $|3  \rangle$ are associated with right, left, up and down  edges respectively. 

\item  $S_{le}$: The shift operator associated with four long-range edges at each vertex is given by
\begin{align} 
S_{le}=&\sum_{v_{y}=1}^{\sqrt{N}}\sum_{i_{1}=0}^{n-2}\sum_{j_{1}=0}^{j_{1_{max}}}\biggl( |5 \rangle \langle 4 |\otimes |i_{1},j_{1}+1;v_{y}\rangle \langle i_{1},j_{1};v_{y}|+
|4 \rangle \langle 5 |\otimes |i_{1},j_{1}-1;v_{y}\rangle \langle i_{1},j_{1};v_{y}|\biggr) \notag \\
+&\sum_{v_{x}=1}^{\sqrt{N}}\sum_{i_{2}=0}^{n-2}\sum_{j_{2}=0}^{j_{2_{max}}}\biggl( |7 \rangle \langle 6 |\otimes |v_{x};i_{2},j_{2}+1 \rangle \langle v_{x};i_{2},j_{2}|+
|6 \rangle \langle 7 |\otimes |v_{x};i_{2},j_{2}-1\rangle \langle v_{x};i_{2},j_{2}|\biggr) \notag \\
+&\sum_{v_{y}=1}^{\sqrt{N}}\biggl( |5 \rangle \langle 4 |+
|4 \rangle \langle 5 |\biggr)\otimes \biggr(| \sqrt{N}/2; v_{y}\rangle \langle \sqrt{N}/2; v_{y}|+| \sqrt{N}; v_{y}\rangle \langle \sqrt{N}; v_{y}|\biggr) \notag  \\
+&\sum_{v_{x}=1}^{\sqrt{N}}\biggl( |7 \rangle \langle 6 |+
|6 \rangle \langle 7 |\biggr)\otimes \biggr(|v_{x}; \sqrt{N}/2\rangle \langle v_{x}; \sqrt{N}/2|+|v_{x}; \sqrt{N} \rangle \langle v_{x}; \sqrt{N}|\biggr)
\label{fshift1}
\end{align}
where the coin basis states $|4 \rangle, |5 \rangle, |6  \rangle$ and $|7  \rangle$ are associated with right, left, up and down long range edges respectively. 

\item $S_{lp}$: Finally the shift operator associated with the undirected self-loop of lackadaisical quantum walk is given by 
\begin{align}
    S_{lp}=&\sum_{v_{y}=1}^{\sqrt{N}} \sum_{v_{x}=1}^{\sqrt{N}} | \mbox{lp}\rangle \langle  \mbox{lp}| \otimes |v_{x};v_{y}\rangle \langle v_{x};v_{y}|.
\end{align}
\end{enumerate}
The action of  the above  flip-flop shift operator on the  tensor product of coin  and vertex space can be summarized as 
\begin{eqnarray}
S| 0 \rangle \otimes | v_x; v_y \rangle &=&  S_{se}| 0 \rangle \otimes | v_x; v_y \rangle =| 1 \rangle \otimes | v_x + 1; v_y\rangle\,, \\
S| 1 \rangle \otimes | v_x; v_y \rangle &=&  S_{se}| 1 \rangle \otimes | v_x; v_y \rangle   = | 0 \rangle \otimes | v_x - 1; v_y\rangle\,, \\
S| 2 \rangle \otimes | v_x; v_y \rangle  &=&  S_{se}| 2 \rangle \otimes | v_x; v_y \rangle  = | 3 \rangle \otimes | v_x; v_y + 1\rangle\,, \\
S| 3 \rangle \otimes | v_x; v_y \rangle  &=&  S_{se}| 3 \rangle \otimes | v_x; v_y \rangle = | 2 \rangle \otimes | v_x; v_y - 1\rangle\,, \\
S| 4 \rangle \otimes | i_{1},j_{1}; v_{y} \rangle &=& S_{le}| 4 \rangle \otimes | i_{1},j_{1}; v_{y} \rangle = | 5 \rangle \otimes | i_{1},j_{1}+1; v_{y}\rangle\,, \\
S| 5 \rangle \otimes | i_{1},j_{1}; v_{y} \rangle  &=&  S_{le}| 5 \rangle \otimes | i_{1},j_{1}; v_{y} \rangle  = | 4 \rangle \otimes | i_{1},j_{1}-1; v_{y}\rangle\,, \\
S| 6 \rangle \otimes | v_x; i_{2},j_{2} \rangle &=&  S_{le}| 6 \rangle \otimes | v_x; i_{2},j_{2} \rangle = | 7 \rangle \otimes | v_x; i_{2},j_{2}+1\rangle\,,  \\
S| 7 \rangle \otimes |  v_; i_{2},j_{2} \rangle  &=& S_{le}| 7 \rangle \otimes |  v_; i_{2},j_{2} \rangle =  | 6 \rangle \otimes |  v_x; i_{2},j_{2}-1\rangle\,, \\
S| \mbox{lp} \rangle \otimes | v_x; v_y \rangle &=& S_{lp}| \mbox{lp} \rangle \otimes | v_x; v_y \rangle = | \mbox{lp} \rangle \otimes | v_x; v_y \rangle\,.
\label{2Dcgrov}
\end{eqnarray}
The success probability to find the marked vertices becomes
\begin{align}
 p_{2d}(t)=\sum_{i=1}^{M}\Bigl(\sum^{7}_{{v_c} = 0 }|\langle v_c | \otimes |\langle t_{i}|\mathcal{U}^{t}|\psi_{2d}(0) \rangle |^{2}+ |\langle  \mbox{lp} | \otimes \langle t_{i}|  \mathcal{U}^{t}|\psi_{2d}(0) \rangle |^{2}\Bigr),
\end{align}
where sum over coin basis states is assumed to   done at the marked vertices.

\subsection{Non-exceptional configurations for the two-dimensional periodic lattice with extra long-range edges} 
In the two-dimensional periodic lattice with extra long-range  edges, there are characteristic non-exceptional configurations. First configuration is non-adjacent pair of vertices. This configuration is not exceptional configuration even on the  two-dimensional periodic lattice without HN4 long-range edges. Second configuration is diagonal vertices, which is exceptional in the case of two-dimensional lattice. However,  from our experiment on two-dimensional lattice  with extra long-range edges, we observe that  this diagonal configuration becomes  non-exceptional configuration.

\subsubsection{Non-adjacent pair of vertices}
We consider pair marked vertices that are not adjacent  with respect to standard  edges and  also with respect to long-range  edges. These  configurations   are non-exceptional configurations.
\begin{figure}[h!]
    \centering
    \includegraphics[scale=0.60]{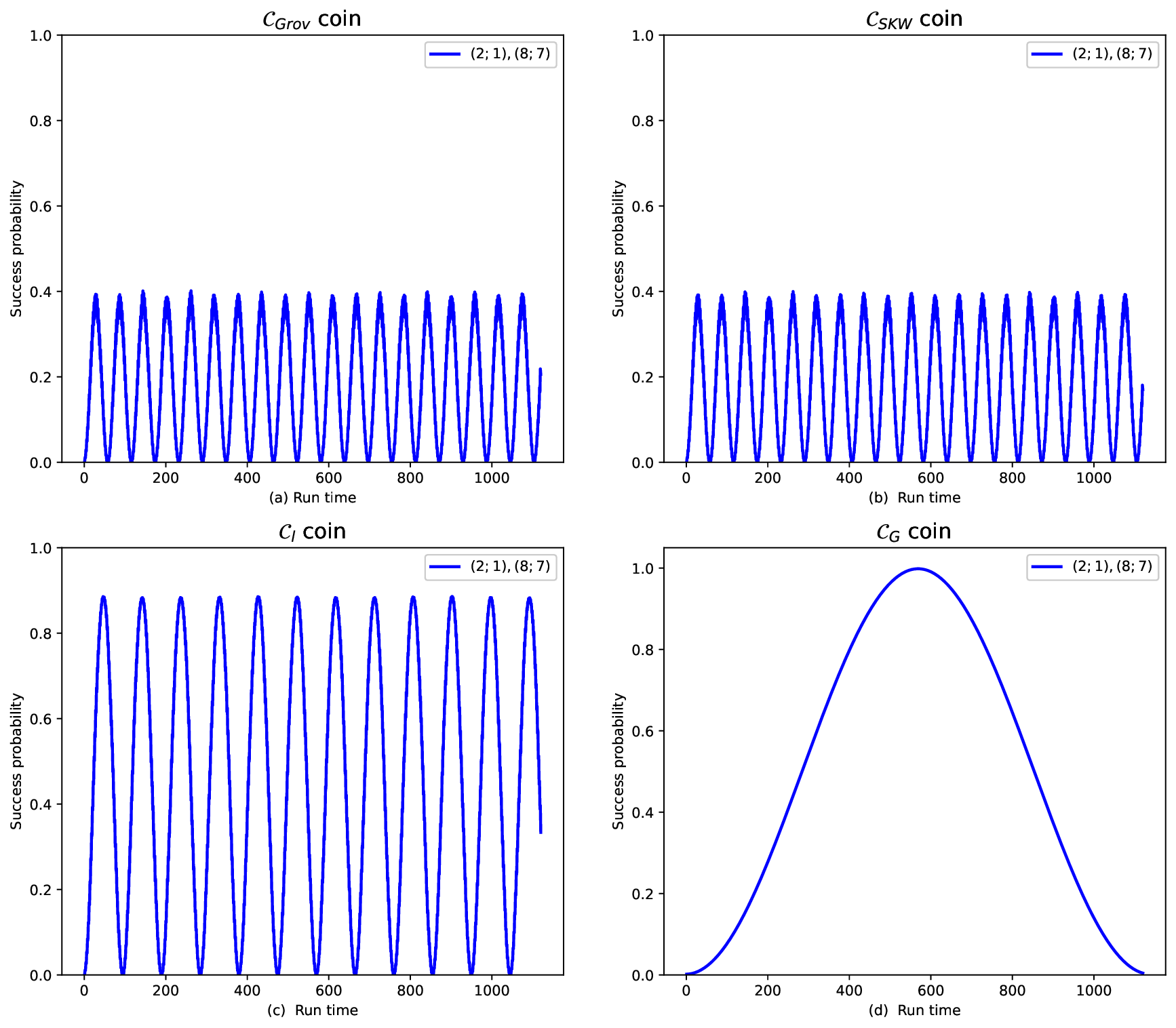}
 \caption{Success probability to measure  two non-adjacent vertices   $(2;1), (8;7)$ on  $2^{5} \times 2^{5}$  two-dimensional periodic lattice with extra long-range  edges as a function of the number of iteration steps for (a) $\mathcal{C}_{Grov}$, (b) $\mathcal{C}_{SKW}$, (c) $\mathcal{C}_{l}$, and (d) $\mathcal{C}_{G}$ coins.}
  \label{2D-nonadjacent}
\end{figure}

{\it{Experimental results}}: The behavior of the success probability to measure two marked vertices, which are non-adjacent,  as a function of the number of iteration steps has been presented in FIG. \ref{2D-nonadjacent}. In this experiment, we take two vertices  $(2;1)$ and $(8;7)$ on a $2^{5} \times 2^{5}$ lattice.  We fix the self-loop weight at $l=8.0/N$. Usually,  we have to tune the self-loop parameter $l$ in a case by  case basis,  such that the success probability is maximized. 
We see that the success probability for the four coins grow as a function of the number of iteration steps. For the Grover and SKW coins in FIGs. \ref{2D-nonadjacent}(a) and \ref{2D-nonadjacent}(b) respectively,  the success probability is not significantly high, so we need to  apply the  amplitude amplification  technique to further  enhance the success probability. 

\subsubsection{Diagonal configuration}
In diagonal  configuration, marked vertices are $(1;1), (2;2),\cdots, (\sqrt{N};\sqrt{N})$. In two-dimensional periodic lattice  without  extra long-range edges, this  is exceptional configuration \cite{amba5}. It has been shown in ref. \cite{men} that this diagonal configuration is a special case of a more general exceptional configuration of the diagonal type.
\begin{figure}[h!]
    \centering
    \includegraphics[scale=0.60]{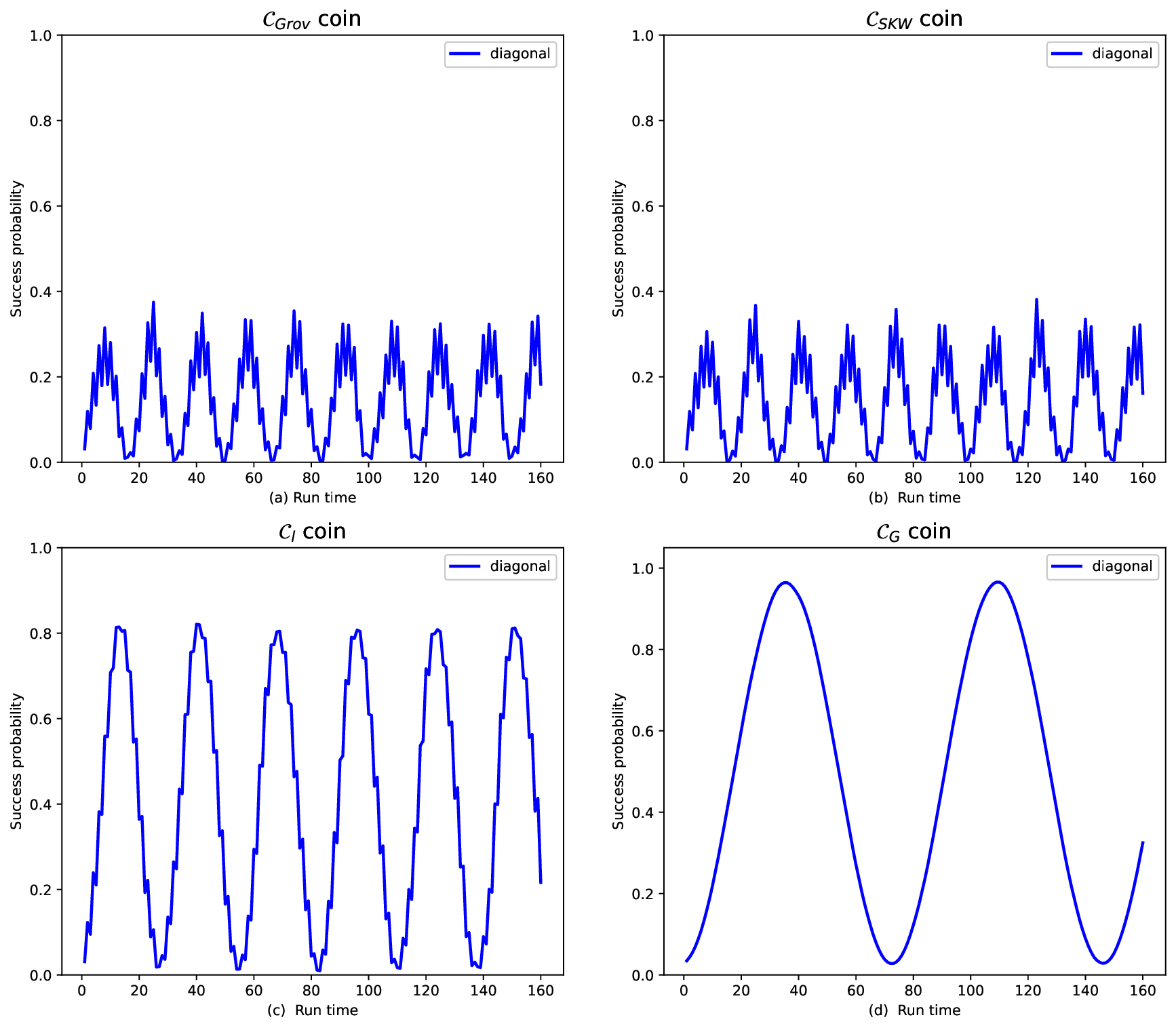}
  \caption{Success probability to measure the diagonal configuration on $2^{5} \times 2^{5}$ two-dimensional periodic lattice with extra long-range  edges as a function of the number of iteration steps for (a) $\mathcal{C}_{Grov}$, (b) $\mathcal{C}_{SKW}$, (c) $\mathcal{C}_{l}$,  and (d) $\mathcal{C}_{G}$ coins.}
  \label{2D-diagonal}
\end{figure}

{\it Experimental results:}  The behavior of the success probability to measure diagonal configuration as a function of the number of iteration steps has been presented in FIG.\ref{2D-diagonal}.  In this experiment, we take diagonal marked vertices  $(1;1),(2;2),\cdots,(32;32)$ on a  $2^{5} \times 2^{5}$ lattice. We fix  the self-loop weight at $l=128.0/N$. 
We see that the success probability for the four coins grow as a function of the number of iteration steps. For the Grover  and SKW coins in FIGs.  \ref{2D-diagonal}(a) and \ref{2D-diagonal}(b) respectively, the success probability is not significantly high, so we need to  apply amplitude amplification to further enhance the success probability.  In case of  two-dimensional periodic lattice, there are special states associated with the diagonal vertices \cite{giri4}. However,  with extra long-range edges, there are no such  special state associated with diagonal vertices. So we can search diagonal configuration by all the  four coins.

\subsection{Exceptional configurations  for the two-dimensional periodic lattice with extra long-range  edge} 
In this subsection, we investigate three types of exceptional configurations. First case is two adjacent marked vertices associated with the  standard   edges. This configuration is also exceptional configuration even  on  two-dimensional periodic lattice without long-range edges.  Second case is two  adjacent marked vertices associated with the long-range  edges. This configuration arises due to the long-range edges. Third case is  the marked vertices associated with directed self-loop formed by some of the long-range edges. Similar  configuration is already studied in one-dimensional case.

\subsubsection{Searching pair of  vertices  adjacent with respect to  standard edges}

\begin{figure}[h!]
    \centering
    \includegraphics[scale=0.60]{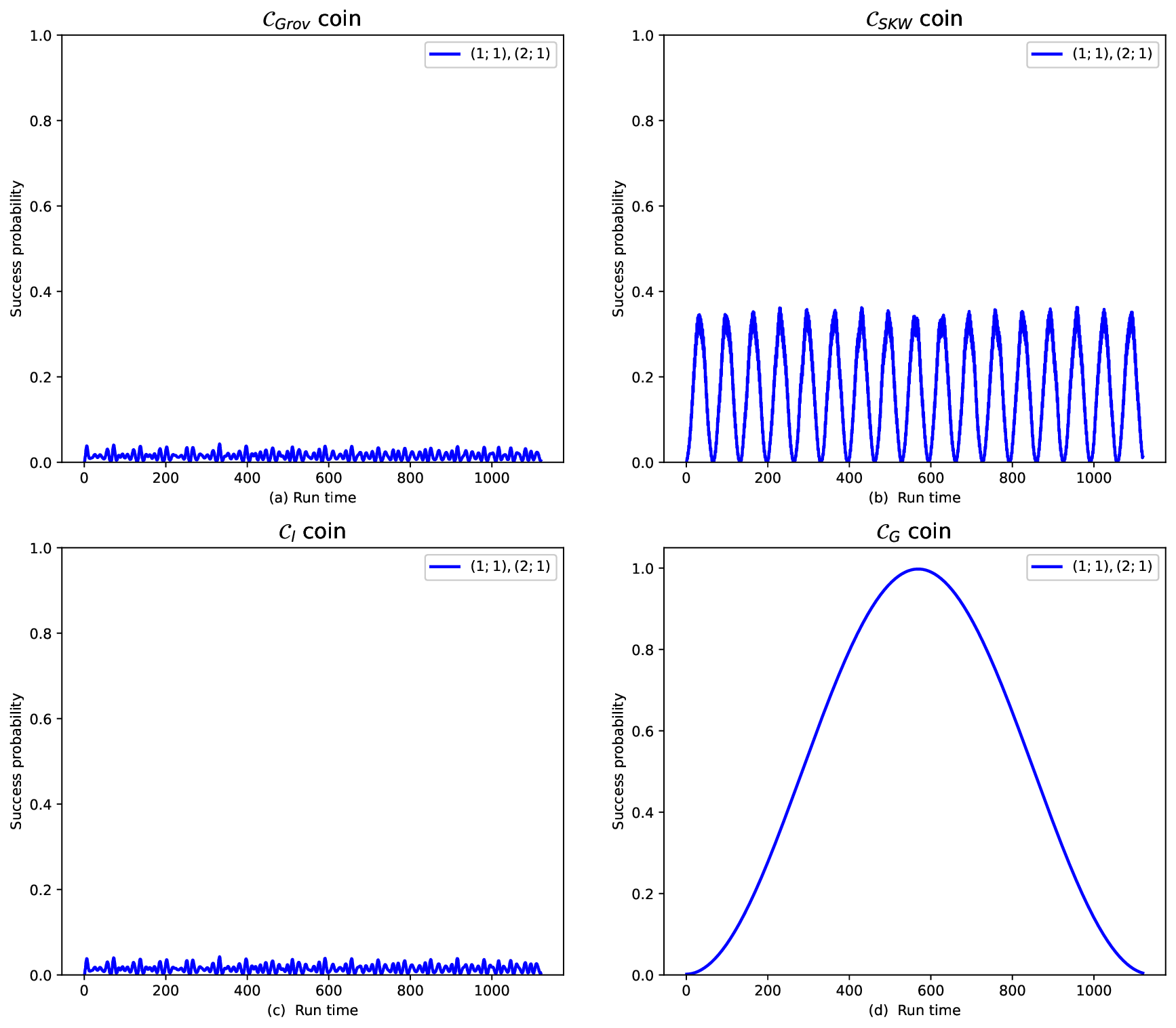}
  \caption{Success probability to measure  two adjacent marked vertices  connected  by standard edges $(1;1),(2;1)$ on a  $2^5 \times 2^5$ two-dimensional periodic lattice with extra long-range  edges as a function of the number of iteration steps for (a) $\mathcal{C}_{Grov}$, (b) $\mathcal{C}_{SKW}$, (c) $\mathcal{C}_{l}$, and (d) $\mathcal{C}_{G}$ coins.}
  \label{2D-short}
\end{figure}

{\it Experimental results:} The behavior of the success probability to measure two adjacent marked vertices  connected by standard  edges  $(1;1), (2;1)$ on a $2^5 \times 2^5$ lattice as a function of the number of iteration steps has been presented in FIG. \ref{2D-short}. We fix self-loop parameter  at $l= 8.0/N$.  
We see that the success probability for $\mathcal{C}_{Grov}$ and $\mathcal{C}_{l}$ coins  represented by FIGs. \ref{2D-short}(a) and \ref{2D-short}(c)  respectively  do not grow at all even if we increase the number of iteration steps. For the $\mathcal{C}_{SKW}$ and $\mathcal{C}_{G}$ coins success probability grow as a function of the number of iteration steps represented by \ref{2D-short}(b) and \ref{2D-short}(d) respectively. For the SKW coin, as we can see, the success probability is not significantly high, so we need to  apply amplitude amplification to further  enhance the success probability. The experiments are carried out for $(1;1), (2;1)$ and $(3;1), (4;1)$. The behavior of the success probability to measure another  two adjacent marked vertices $(3;1), (4;1)$,  that are connected by standard  edges  on the  $2^5 \times 2^5$ lattice as a function of the number of iteration steps is also  similar to FIG. \ref{2D-short}. The $x$ coordinate and the $y$ coordinate of $(1;1)=(0,0;0;0)$ are $i=0$ hierarchy of HN4. The $x$ coordinate of $(2;1)=(1,0;0;0)$ is $i=1$ hierarchy of HN4 and the $y$ coordinate of $(2;1)=(1,0;0;0)$ is $i=0$ hierarchy of HN4. Whereas The $x$ coordinate and the $y$ coordinate of $(3;1)=(0,1;0;0)$ are $i=0$ hierarchy of HN4. The $x$ coordinate of $(4;1)=(2,0;0;0)$ is $i=2$ hierarchy of HN4 and the $y$ coordinate of $(4;1)=(0,1;0;0)$ is $i=0$ hierarchy of HN4. This lattice doesn't have translation invariance, so success probabilities for two pairs  $(1;1), (2;1)$ and $(3;1), (4;1)$ are slight different but very similar. Experimental result for the other adjacent vertices also looks very similar.  The reason that these configurations cannot be searched by $\mathcal{C}_{Grov}$ and $\mathcal{C}_{l}$ coins is explained theoretically by stationary states.

Similar to the one-dimensional case,  the stationary state here exhibits substantial overlap with the initial state of the quantum search.  In this sub-subsection, we focus on constructing the stationary state for the exceptional configuration arising from two adjacent vertices with respect to the standard edges. 
The stationary state is defined as the eigenvector of the time evolution operator with a unit eigenvalue. This vector is crucial for understanding why  $\mathcal{C}_{Grov}$ and  $\mathcal{C}_{l}$ coins are ineffective at searching for these type of exceptional configurations. To describe the action of the coin operators on the stationary state,  the following pair of non-normalized coin states
\begin{eqnarray}
|\psi_{c}^{\mathrm{se},x+} \rangle =|\psi_{cl}(0) \rangle  - \sqrt{8+l}|0 \rangle, 
~~~|\psi_{c}^{\mathrm{se},x-} \rangle =|\psi_{cl}(0) \rangle  - \sqrt{8+l}|1 \rangle\,,
\end{eqnarray}
and 
\begin{eqnarray}
|\psi_{c}^{\mathrm{se},y+} \rangle =|\psi_{cl}(0) \rangle  - \sqrt{8+l}|2 \rangle, 
~~~|\psi_{c}^{\mathrm{se},y-} \rangle =|\psi_{cl}(0) \rangle  - \sqrt{8+l}|3 \rangle\,,
\end{eqnarray}
are very much useful. 
Note that, the above four states are orthogonal to $|\psi_{cl}(0) \rangle $, i.e.\; $\langle \psi_{c}^{\mathrm{se},x+}|\psi_{cl}(0) \rangle=\langle \psi_{c}^{\mathrm{se},x-}|\psi_{cl}(0) \rangle=\langle \psi_{c}^{\mathrm{se},y+}|\psi_{cl}(0) \rangle=\langle \psi_{c}^{\mathrm{se},y-}|\psi_{cl}(0) \rangle=0$.
First,  we consider the stationary state associated with a pair of  horizontal  marked vertices adjacent  with respect to the standard edges. This state is written by all the unmarked vertices with their coin state of the form $|\psi_{cl}(0) \rangle$ and the two marked vertices $(v_{x};v_{y})$ and $(v_{x}+1;v_{y})$ with their coin states of the form $|\psi_{c}^{\mathrm{se},x+} \rangle $ and $|\psi_{c}^{\mathrm{se},x-} \rangle $ respectively.
The stationary state associated with two horizontal marked vertices  adjacent with respect to the  standard  edges is defined as follows 
\begin{align}
|\psi_{hstat,se} \rangle =|\psi_{2d}(0) \rangle - \sqrt{\frac{8+l}{N}} \biggl( |0 \rangle \otimes |v_{x};v_{y} \rangle  + |1 \rangle \otimes |v_{x}+1;v_{y} \rangle \biggr) ,
\end{align}
where the initial state $|\psi_{2d}(0)$  is given by eq. (\ref{in2dl}).
Next,  we consider the stationary state associated with two vertical marked vertices,  adjacent with respect to the  standard  edges.
This state is defined as follows
\begin{align}
|\psi_{vstat,se} \rangle =|\psi_{2d}(0) \rangle - \sqrt{\frac{8+l}{N}} \biggl( |2 \rangle \otimes |v_{x};v_{y} \rangle  + |3  \rangle \otimes |v_{x};v_{y}+1 \rangle \biggr) .
\end{align}
Bellow is the action of the four coin operators on these  stationary state:
\begin{enumerate}
\item $U_{l}$: We can demonstrate  that the stationary states $|\psi_{hstat,short} \rangle$ and $|\psi_{vstat,short} \rangle$ are eigenstates of the time evolution operator  $U_{l}=S\mathcal{C}_{l}$  with a unit eigenvalue
\begin{align}
    U_{l} |\psi_{hstat,se} \rangle=|\psi_{hstat,se} \rangle, \\
    U_{l} |\psi_{vstat,se} \rangle=|\psi_{vstat,se} \rangle.
\end{align}

\item $U_{Grov}$: Similarly,  $|\psi_{hstat,se} \rangle$ and $|\psi_{vstat,se} \rangle$ remain  eigenstates for  $U_{Grov}= S\mathcal{C}_{Grov} =  U_{l=0}$: 
\begin{align}
    U_{Grov} |\psi_{hstat,se} \rangle =|\psi_{hstat,se} \rangle, \\
    U_{Grov} |\psi_{vstat,se} \rangle =|\psi_{vstat,se} \rangle.
\end{align}
Here we have taken  $l=0$ limit, since quantum walk with Grover coin does not have self-loop.  
\item $U_{G}$: However, under the operator $U_{G}=S\mathcal{C}_{G}$, the stationary states do  not remain stationary, as illustrated by the following expressions 
\begin{align}
    U_{G} |\psi_{hstat,se} \rangle &=|\psi_{hstat,se} \rangle-\frac{2}{\sqrt{N}}S\biggl(|\psi_{c}^{\mathrm{se},x+} \rangle \otimes |v_{x};v_{y} \rangle +|\psi_{c}^{\mathrm{se},x-} \rangle \otimes |v_{x}+1;v_{y} \rangle \biggr)  \nonumber \\
    &- 2\sqrt{\frac{l}{N(8+l)}}SD_{l}\Biggl[|\mbox{lp} \rangle \otimes \biggl( |v_{x};v_{y} \rangle +|v_{x}+1;v_{y} \rangle  \biggr)\Biggl],\\
    U_{G} |\psi_{vstat,se} \rangle &=|\psi_{vstat,se} \rangle-\frac{2}{\sqrt{N}}S\biggl(|\psi_{c}^{\mathrm{se},y+} \rangle \otimes |v_{x};v_{y} \rangle +|\psi_{c}^{\mathrm{se},y-} \rangle \otimes |v_{x};v_{y}+1 \rangle \biggr)  \nonumber \\
    &- 2\sqrt{\frac{l}{N(8+l)}}SD_{l}\Biggl[|\mbox{lp} \rangle \otimes \biggl( |v_{x};v_{y} \rangle +|v_{x};v_{y}+1 \rangle  \biggr)\Biggl],
\end{align}

\item $U_{SKW}$: Similarly, under $U_{SKW}=S \mathcal{C}_{SKW}$, the stationary states also do not remain stationary, as shown by the following expressions 
\begin{align}
    U_{SKW} |\psi_{hstat,se}  \rangle =|\psi_{hstat,se} \rangle-\frac{2}{\sqrt{N}}S\biggl(|\psi_{c}^{\mathrm{se},x+} \rangle \otimes |v_{x};v_{y} \rangle +|\psi_{c}^{\mathrm{se},x-} \rangle \otimes |v_{x}+1;v_{y} \rangle \biggr), \\
    U_{SKW} |\psi_{vstat,se}  \rangle =|\psi_{vstat,se} \rangle-\frac{2}{\sqrt{N}}S\biggl(|\psi_{c}^{\mathrm{se},y+} \rangle \otimes |v_{x};v_{y} \rangle +|\psi_{c}^{\mathrm{se},y-} \rangle \otimes |v_{x};v_{y}+1 \rangle \biggr).
\end{align}
\end{enumerate}
Here also we take $l=0$ limit, because SKW coin does not have self-loop. 
We can express the initial state $|\psi_{2d}(0)\rangle $ in terms of the stationary state $|\psi_{hstat,se} \rangle $ as
\begin{align}
|\psi_{2d}(0) \rangle =|\psi_{hstat,se} \rangle + \sqrt{\frac{8+l}{N}} \biggl( |0 \rangle \otimes |v_{x};v_{y} \rangle  + |1 \rangle \otimes |v_{x}+1;v_{y} \rangle \biggr)
\label{initialof2dshort}
\end{align}
The final state, after applying $U_{l}$ repeatedly $t$ times, becomes 
\begin{align}
U_{l}^{t}|\psi_{2d}(0) \rangle  =|\psi_{hstat,se} \rangle + \sqrt{\frac{8+l}{N}} U_{l}^{t}\biggl( |0  \rangle \otimes |v_{x};v_{y} \rangle  + |1 \rangle \otimes |v_{x}+1;v_{y} \rangle \biggr).
\label{time2dshort}
\end{align}
It is important to note that the first component on the right side of eq.  \eqref{time2dshort} remains unchanged under the action of the evolution operator; only the second component is affected. The upper bound for the success probability to find two  adjacent vertices with respect to the  standard  edges can be determined by setting $U_{l}^{t} = -\mathbb{I}$ in the second term. Consequently, the success probability for the  two marked vertices that are adjacent with respect to the  standard  edges is constrained by  $p_{2d}(t) \le \mathcal{O}(1/N)$. The same arqument also applies  to the case of $U_{Grov}$. It indicates that the quantum walk search  conducted using $U_{l}$ and $U_{Grov}$ on a two-dimensional periodic lattice with extra long-range  edges are unable to find two  marked edges  that are adjacent with respect to the  standard  edges, as the success probability remains limited by the initial success probability. 
However, when $U_{G}$ is applied to the initial state, both the first and second components of eq.  \eqref{initialof2dshort} undergo non-trivial transformations, since $|\psi_{hstat,se} \rangle$ is no longer an eigenstate of $U_{G}$. Following repeated applications of the evolution operator, the final state achieves a significantly high and constant  overlap with the marked vertices.
In the case of $U_{SKW}$, both components of the initial state also transform non-trivially. However, to achieve a high success probability, it is necessary to implement amplitude amplification in conjunction with the repeated application of $U_{SKW}$ to the initial state.  

\subsubsection{Searching pair of  vertices  adjacent with respect to  long-range edges}

\begin{figure}[h!]
    \centering
    \includegraphics[scale=0.60]{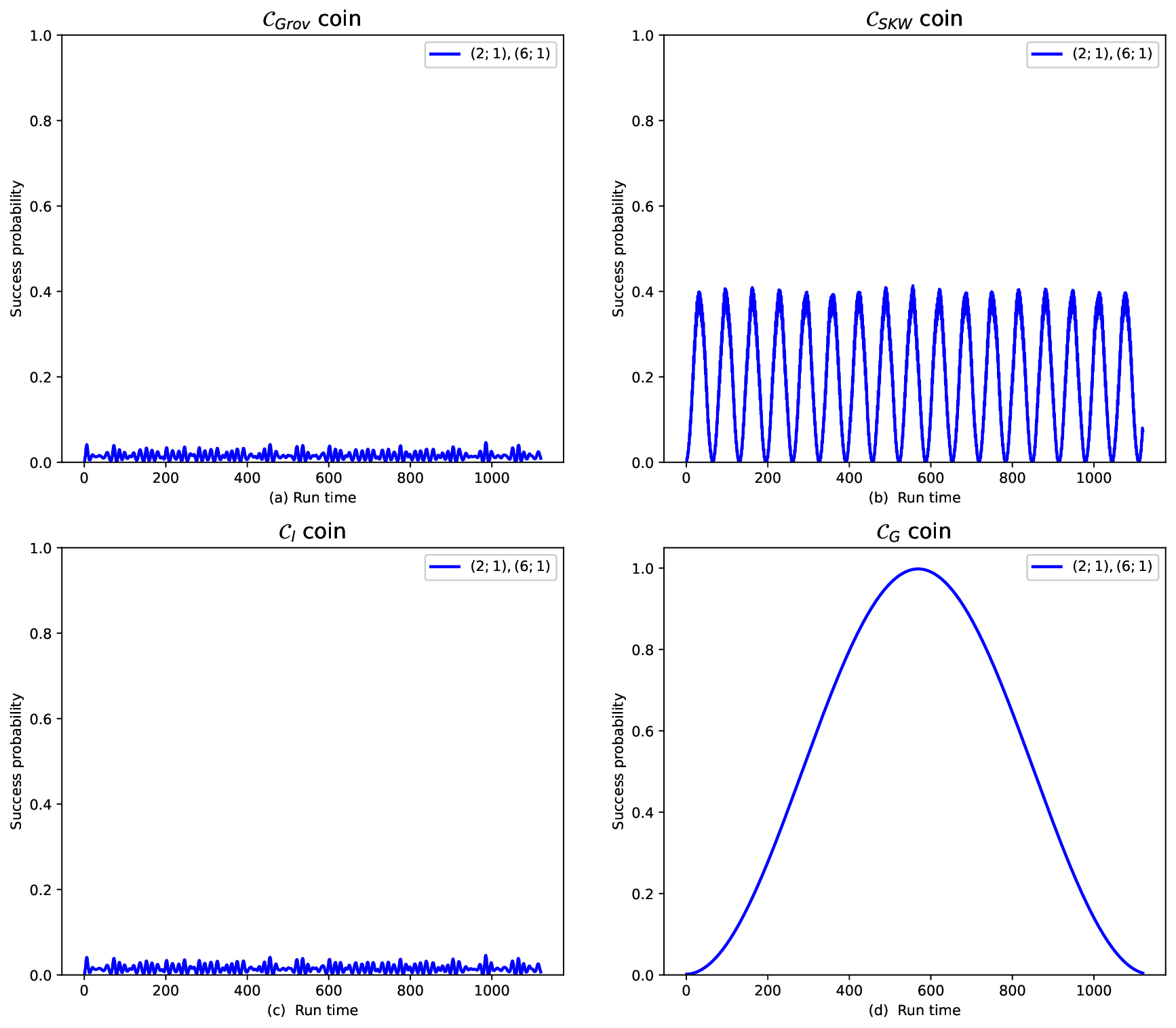}
    \caption{Success probability to measure  two marked vertices $(2;1),(6;1)$, adjacent with respect  to  the long-range  edges on  a  $2^5 \times 2^5$ two-dimensional periodic lattice with extra long-range  edges as a function of the number of iteration steps for (a) $\mathcal{C}_{Grov}$, (b) $\mathcal{C}_{SKW}$, (c) $\mathcal{C}_{l}$, and (d) $\mathcal{C}_{G}$ coins.}
  \label{2D-long}
\end{figure}

{\it Experimental results:} The behavior of the success probability to measure two marked vertices $(2;1), (6;1)$, adjacent with respect to the  long-range edges  on a $2^5 \times 2^5$ lattice as a function of the number of iteration steps has been presented in FIG.  \ref{2D-long}. We fix the  self-loop parameter at fixed value $l= 8.0/N$.    We see that the success probability for  $\mathcal{C}_{Grov}$ and $\mathcal{C}_{l}$ coins  represented in FIGs.  \ref{2D-long}(a) and \ref{2D-long}(c) respectively,  do not grow at all even if we increase the number of iteration steps. For $\mathcal{C}_{SKW}$ and $\mathcal{C}_{G}$ coins  the success probability grow as a function of the number of iteration steps represented by FIGs. \ref{2D-long}(b) and \ref{2D-long}(d) respectively. For the SKW coin, as we can see, the success probability is not significantly high, so we need to  apply amplitude amplification to further enhance the success probability. The result is reported only  for $(2;1), (6;1)$ vertices, however, the results for other pair of such vertices  are similar.  The reason that these configurations cannot be searched by $\mathcal{C}_{Grov}$ and $\mathcal{C}_{l}$ coins can be  explained in a similar fashion like the previous subsection.

In this sub-subsection, we construct the stationary states  to the exceptional configuration occurred from  two marked vertices,  adjacent with respect to the long-range  edge. The stationary state is the eigenvector of the time evolution operator with unit eigenvalue. For the description of the  action of evolution operators on  the stationary state, we need the following pair of non-normalized coin states
\begin{eqnarray}
|\psi_{c}^{\mathrm{le},x+} \rangle =|\psi_{cl}(0) \rangle  - \sqrt{8+l}|4 \rangle, 
~~~|\psi_{c}^{\mathrm{le},x-} \rangle =|\psi_{cl}(0) \rangle  - \sqrt{8+l}|5 \rangle\,,
\end{eqnarray}
and
\begin{eqnarray}
|\psi_{c}^{\mathrm{le},y+} \rangle =|\psi_{cl}(0) \rangle  - \sqrt{8+l}|6 \rangle,
~~~|\psi_{c}^{\mathrm{le},y-} \rangle =|\psi_{cl}(0) \rangle  - \sqrt{8+l}|7 \rangle.
\end{eqnarray}
Note that, the above four states are orthogonal to $|\psi_{cl}(0) \rangle $, i.e.\; $\langle \psi_{c}^{\mathrm{le},x+}|\psi_{cl}(0) \rangle=\langle \psi_{c}^{\mathrm{le},x-}|\psi_{cl}(0) \rangle=\langle \psi_{c}^{\mathrm{le},y+}|\psi_{cl}(0) \rangle=\langle \psi_{c}^{\mathrm{le},y-}|\psi_{cl}(0) \rangle=0$.
First,  we consider the stationary state associated with two horizontal marked vertices adjacent with respect to the long-range  edges. This state is written by all the unmarked vertices with their coin states  of the form $|\psi_{cl} \rangle$ and the two marked vertices $(i_{1},j_{1};,v_{y})$ and $(i_{1},j_{1}+1;v_{y})$ with their coin states of the form $|\psi_{c}^{\mathrm{le},x+} \rangle $ and $|\psi_{c}^{\mathrm{le},x-} \rangle $ respectively.
The stationary state associated with these two horizontal adjacent marked vertices  is defined as follows
\begin{align}
|\psi_{hstat,le} \rangle =|\psi_{2d}(0) \rangle - \sqrt{\frac{8+l}{N}} \biggl( |4  \rangle \otimes |i_{1},j_{1};v_{y} \rangle  + |5  \rangle \otimes |i_{1},j_{1}+1;v_{y} \rangle \biggr)\,.
\end{align}
Second, we consider the stationary state associated with two vertical marked vertices, adjacent with respect to the  long-range edges.  This state is defined as follows 
\begin{align}
|\psi_{vstat,le} \rangle =|\psi_{2d}(0) \rangle - \sqrt{\frac{8+l}{N}} \biggl( |6  \rangle \otimes |v_{x};i_{2},j_{2} \rangle  + |7  \rangle \otimes |v_{x};i_{2},j_{2}+1 \rangle \biggr) .
\end{align}
Bellow is the action of the four coin operators on these  stationary state:
\begin{enumerate}
\item $U_{l}$: We can demonstrate  that the stationary states $|\psi_{hstat,le} \rangle$ and $|\psi_{vstat,le} \rangle$  are eigenstates of the time evolution operator  $U_{l}=S\mathcal{C}_{l}$, with a unit eigenvalue
\begin{align}
    U_{l} |\psi_{hstat,le} \rangle=|\psi_{hstat,le} \rangle, \\
    U_{l} |\psi_{vstat,le} \rangle=|\psi_{vstat,le} \rangle.
\end{align}

\item $U_{Grov}$: $|\psi_{hstat,le} \rangle$ and $|\psi_{vstat,le} \rangle$ remain eigenstates for the evolution operator $U_{Grov}= S\mathcal{C}_{Grov}= U_{l=0}$ with Grover coin 
\begin{align}
    U_{Grov} |\psi_{hstat,le} \rangle =|\psi_{hstat,le} \rangle, \\
    U_{Grov} |\psi_{vstat,le} \rangle=|\psi_{vstat,le} \rangle.
\end{align}
Here we take $l=0$ limit, since Grover coin with regular quantum walk does not have self-loop.  

\item $U_{G}$: However, under the operator $U_{G}=S\mathcal{C}_{G}$, the stationary states do not remain stationary, as illustrated by the following expressions 
\begin{align}
    U_{G} |\psi_{hstat,le} \rangle &=|\psi_{hstat,le} \rangle-\frac{2}{\sqrt{N}}S\biggl(|\psi_{c}^{\mathrm{le},x+} \rangle \otimes |i_{1},j_{1};v_{y} \rangle +|\psi_{c}^{\mathrm{le},x-} \rangle \otimes |i_{1},j_{1}+1;v_{y} \rangle \biggr)  \nonumber \\
    &- 2\sqrt{\frac{l}{N(8+l)}}SD_{l}\Biggl[|\mbox{lp} \rangle \otimes \biggl( |i_{1},j_{1};v_{y} \rangle +|i_{1},j_{1}+1;v_{y} \rangle  \biggr)\Biggl],\\
    U_{G} |\psi_{vstat,le} \rangle &=|\psi_{vstat,le} \rangle-\frac{2}{\sqrt{N}}S\biggl(|\psi_{c}^{\mathrm{le},y+} \rangle \otimes |v_{x};i_{2},j_{2} \rangle +|\psi_{c}^{\mathrm{le},y-} \rangle \otimes |v_{x};i_{2},j_{2}+1 \rangle \biggr)  \nonumber \\
    &- 2\sqrt{\frac{l}{N(8+l)}}SD_{l}\Biggl[|\mbox{lp} \rangle \otimes \biggl( |v_{x};i_{2},j_{2} \rangle +|v_{x};i_{2},j_{2}+1 \rangle  \biggr)\Biggl],
\end{align}

\item $U_{SKW}$: Similarly, under $U_{SKW}=S \mathcal{C}_{SKW}$, the stationary states also do not remain stationary, as shown by the following expressions 
\begin{align}
    U_{SKW} |\psi_{hstat,le}  \rangle =|\psi_{hstat,le} \rangle-\frac{2}{\sqrt{N}}S\biggl(|\psi_{c}^{\mathrm{le},x+} \rangle \otimes |i_{1},j_{1};v_{y} \rangle +|\psi_{c}^{\mathrm{le},x-} \rangle \otimes |i_{1},j_{1}+1;v_{y} \rangle \biggr), \\
    U_{SKW} |\psi_{vstat,le}  \rangle =|\psi_{vstat,le} \rangle-\frac{2}{\sqrt{N}}S\biggl(|\psi_{c}^{\mathrm{le},y+} \rangle \otimes |v_{x};i_{2},j_{2} \rangle +|\psi_{c}^{\mathrm{le},y-} \rangle \otimes |v_{x};i_{2},j_{2}+1 \rangle \biggr).
\end{align}
\end{enumerate}
Note that, for the quantum walk search with SKW coin there is no self-loop state, so  we set   $l =0$. 
We can express the initial state $|\psi_{2d}(0)\rangle $ in terms of the stationary state $|\psi_{hstat,le} \rangle $ as
\begin{align}
|\psi_{2d}(0) \rangle =|\psi_{hstat,le} \rangle + \sqrt{\frac{8+l}{N}} \biggl( |4 \rangle \otimes |i_{1},j_{1};v_{y} \rangle  + |5  \rangle \otimes |i_{1},j_{1}+1;v_{y} \rangle \biggr)
\label{initialof2dlong}
\end{align}
The final state, after applying $U_{l}$ repeatedly $t$ times, becomes 
\begin{align}
U_{l}^{t}|\psi_{2d}(0) \rangle  =|\psi_{hstat,le} \rangle + \sqrt{\frac{8+l}{N}} U_{l}^{t}\biggl( |4  \rangle \otimes |i_{1},j_{1};v_{y} \rangle  + |5 \rangle \otimes |i_{1},j_{1}+1;v_{y} \rangle \biggr).
\label{time2dlong}
\end{align}
It is important to note that the first component on the right side of eq.  \eqref{time2dlong}  remains unchanged under the action of the evolution operator; only the second component is affected. The upper bound for the success probability of locating two marked vertices  that are adjacent with respect to the  long-range  edges can be determined by setting $U_{l}^{t} = -\mathbb{I}$ in the second term. Consequently, the success probability to find  two marked vertices that are adjacent with respect to the  long-range  edges is constrained by $p_{2d}(t) \le \mathcal{O}(1/N)$. The same argument also holds for $U_{Grov}$. It indicates that the quantum walk search  conducted using $U_{l}$ and $U_{Grov}$ on a two-dimensional periodic lattice with extra long-range  edges are unable to find the  two marked vertices that are adjacent with respect to the  long-range  edges, as the success probability remains limited by the initial success probability. 
However, when  $U_{G}$ is applied to the initial state, both the first and second components of eq. \eqref{initialof2dlong} undergo non-trivial transformation, since $|\psi_{hstat,le} \rangle$ is no longer an eigenstate of $U_{G}$. Following repeated applications of the evolution operator, the final state achieves a significantly high and constant  overlap with the marked vertex.
In the case of $U_{SKW}$, both components of the initial state also transform non-trivially. However, to achieve a high success probability, it is necessary to implement amplitude amplification in conjunction with the repeated application of $U_{SKW}$.

\subsubsection{Searching  for vertices  with one directed self-loop of the long-range edges}

\begin{figure}[h!]
    \centering
    \includegraphics[scale=0.60]{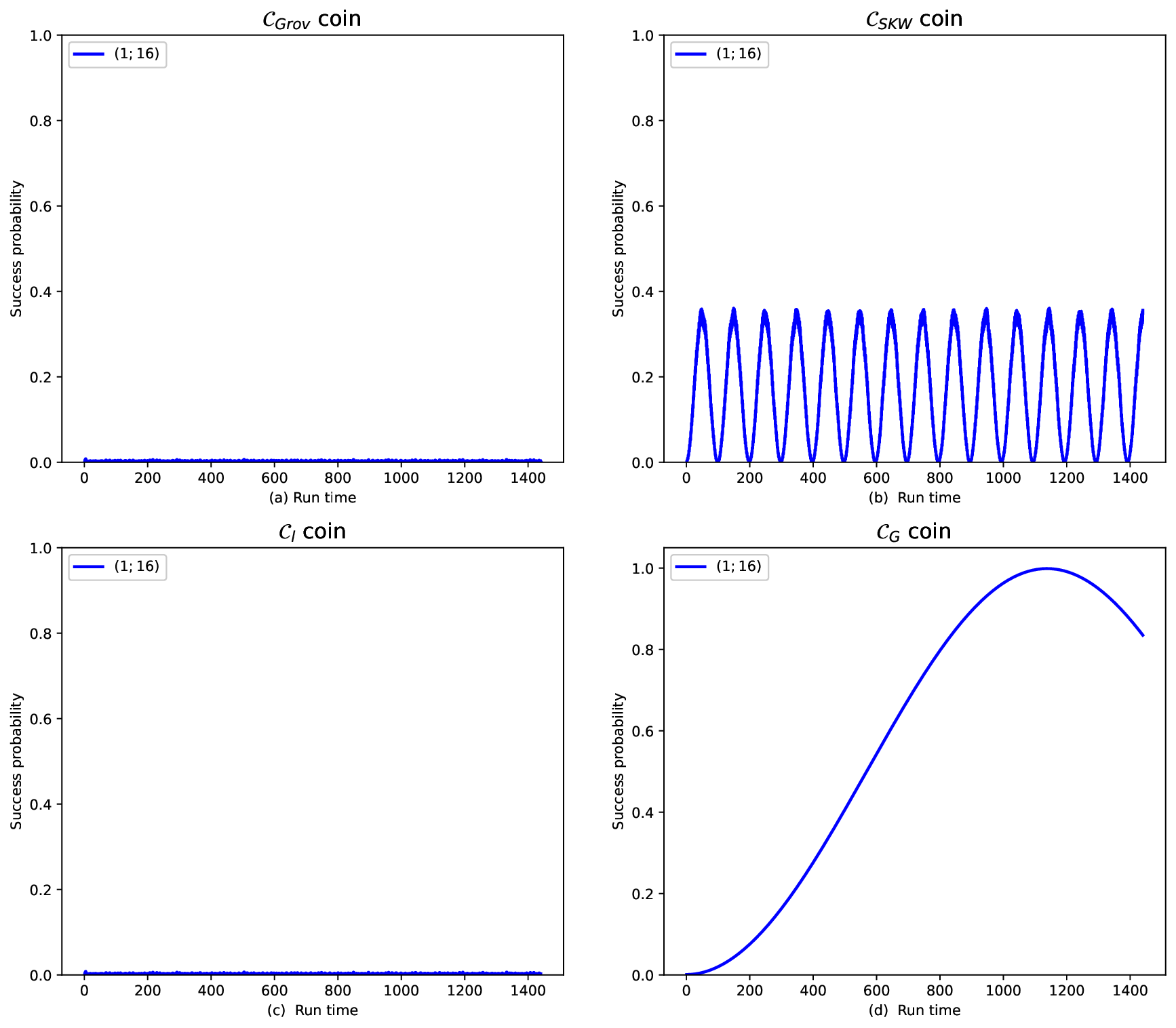}
    \caption{Success probability to measure a marked vertex $(1;16)$ on a  $2^5 \times 2^5$ two-dimensional periodic lattice with extra long-range  edges as a function of the number of iteration steps for (a) $\mathcal{C}_{Grov}$, (b) $\mathcal{C}_{SKW}$, (c) $\mathcal{C}_{l}$, and (d) $\mathcal{C}_{G}$ coins. }
  \label{2D-HN41}
\end{figure}

{\it Experimental results:} The behavior of the success probability to measure the  marked vertex  which has an directed self-loop at $(1;16)$ on a $2^5 \times 2^5$ lattice as a function of the number of iteration steps has been presented in FIG. \ref{2D-HN41}. We fix the self-loop weight at $l= 4.0/N$.   
We see that the success probability for the $\mathcal{C}_{Grov}$ and $\mathcal{C}_{l}$ coins  represented by FIGs. \ref{2D-HN41}(a) and \ref{2D-HN41}(c) respectively  do not grow at all even if we increase the number of iteration steps. For the $\mathcal{C}_{SKW}$ and $\mathcal{C}_{G}$ coins  success probability grow as a function of the number of iteration steps represented by FIGs. \ref{2D-HN41}(b) and \ref{2D-HN41}(d) respectively. For the SKW coin, as we can see, the success probability is not significantly high, so we need to  apply amplitude amplification to further enhance the success probability. 
Although the experimental result  only for $(1;16)$ is plotted in FIG. \ref{2D-HN41},  experimental results  for $(v_{x};16),\;(1 \le v_{x} \le 32, v_{x} \neq 16, 32 )$ are  similar. 
The reason that these configurations cannot be searched by $\mathcal{C}_{Grov}$ and $\mathcal{C}_{l}$ coins  is explained as follows. 
We consider the stationary state, which has large overlap with the  initial state of the quantum walk search. In this sub-subsection, we construct the stationary state of the exceptional configuration occurred due to the formation of directed self-loop by some long-range edges. 
The stationary state is the eigenvector of a time evolution operator with unit eigenvalue. 
For the description of action of  the coin operators on the stationary state, we need the following pairs  of non-normalized coin states $|\psi_{c}^{\mathrm{le},x+} \rangle, |\psi_{c}^{\mathrm{le},x-} \rangle$, and  $|\psi_{c}^{\mathrm{le},y+} \rangle, |\psi_{c}^{\mathrm{le},y-} \rangle$. These states are already defined in  previous subsection.
We here define the stationary states associated with the exceptional vertices  $|\sqrt{N}/2 ;v_{y}\rangle, |\sqrt{N} ;v_{y}\rangle, |v_{x}; \sqrt{N}/2 \rangle$ and $|v_{x}; \sqrt{N} \rangle$. The stationary states associated with $|\sqrt{N}/2 ;v_{y}\rangle$ and $|v_{x}; \sqrt{N}/2 \rangle$ are defined as follow
\begin{align}
|\psi_{HNstat,x} \rangle :=|\psi_{2d}(0) \rangle - \sqrt{\frac{8+l}{N}}\frac{1}{2} \biggl( |4  \rangle +|5  \rangle \biggr)\otimes | \sqrt{N}/2; v_{y} \rangle , \\
|\psi_{HNstat,y} \rangle :=|\psi_{2d}(0) \rangle - \sqrt{\frac{8+l}{N}}\frac{1}{2} \biggl( |6  \rangle +|7  \rangle \biggr)\otimes | v_{x}; \sqrt{N}/2 \rangle  ,
\end{align}
where the initial state $|\psi_{2d}(0)\rangle$  is given by eq. (\ref{in2dl}).  The stationary states associated with $|\sqrt{N}; v_{y}\rangle$ and $|v_{x}; \sqrt{N} \rangle$ can also defined in a similar fashion.
The following  is the action of four coin operators on these stationary states:
\begin{enumerate}
\item $U_{l}$: We can demonstrate that the stationary states $|\psi_{HNstat,x} \rangle$ and $|\psi_{HNstat,y} \rangle$ are eigenstates of the time evolution operator  $U_{l}=S\mathcal{C}_{l}$, with a unit eigenvalue
\begin{align}
    U_{l} |\psi_{HNstat,x} \rangle=|\psi_{HNstat,x} \rangle, \\
    U_{l} |\psi_{HNstat,y} \rangle=|\psi_{HNstat,y} \rangle.
\end{align}

\item $U_{Grov}$: $|\psi_{HNstat,x} \rangle$ and $|\psi_{HNstat,y} \rangle$ remain eigenstates for the evolution operator  $U_{Grov}= S\mathcal{C}_{Grov}=U_{l=0}$ with Grover coin 
\begin{align}
    U_{Grov} |\psi_{HNstat,x} \rangle&=|\psi_{HNstat,x} \rangle, \\
    U_{Grov} |\psi_{HNstat,y} \rangle&=|\psi_{HNstat,y} \rangle\,.
\end{align}
Here we take  $l=0$ limit.

\item $U_{G}$: However, under the operator $U_{G}=S\mathcal{C}_{G}$, the stationary states do not remain stationary, as illustrated by the following expressions
\begin{eqnarray}\nonumber
    U_{G} |\psi_{HNstat,x} \rangle =\hspace{13cm}\\ |\psi_{HNstat,x} \rangle-\frac{1}{\sqrt{N}}S\Biggl[\biggl(|\psi_{c}^{\mathrm{le},x+} \rangle  +|\psi_{c}^{\mathrm{le},x-} \rangle  \biggr) \otimes |\sqrt{N}/2; v_{y} \rangle \Biggl]
    - 2\sqrt{\frac{l}{N(8+l)}}SD_{l}\biggl( |\mbox{lp} \rangle \otimes|\sqrt{N}/2; v_{y} \rangle   \biggr),\\ \nonumber
     U_{G} |\psi_{HNstat,y} \rangle = \hspace{13cm}\\ |\psi_{HNstat,y} \rangle-\frac{1}{\sqrt{N}}S\Biggl[\biggl(|\psi_{c}^{\mathrm{le},y+} \rangle  +|\psi_{c}^{\mathrm{le},y-} \rangle  \biggr) \otimes |v_{x}; \sqrt{N}/2 \rangle \Biggl]  
    - 2\sqrt{\frac{l}{N(8+l)}}SD_{l}\biggl( |\mbox{lp} \rangle \otimes|v_{x}; \sqrt{N}/2 \rangle   \biggr).
\end{eqnarray}

\item $U_{SKW}$: Similarly, under $U_{SKW}=S \mathcal{C}_{SKW}$, the stationary states also do not remain stationary, as shown by the following expressions 
\begin{align}
    U_{SKW} |\psi_{HNstat,x} \rangle   =|\psi_{HNstat,x} \rangle  -\frac{1}{\sqrt{N}}S\Biggl[\biggl(|\psi_{c}^{\mathrm{le},x+} \rangle +|\psi_{c}^{\mathrm{le},x-} \rangle \biggr)\otimes |\sqrt{N}/2; v_{y} \rangle \Biggl],\\
    U_{SKW} |\psi_{HNstat,y} \rangle   =|\psi_{HNstat,y} \rangle  -\frac{1}{\sqrt{N}}S\Biggl[\biggl(|\psi_{c}^{\mathrm{le},y+} \rangle +|\psi_{c}^{\mathrm{le},y-} \rangle \biggr)\otimes |v_{x}; \sqrt{N}/2 \rangle \Biggl].
\end{align}
\end{enumerate}
Note that, for the quantum walk search with SKW coin there is no self-loop, so we  set $l =0$ wherever necessary. 
We can express the initial state $|\psi_{2d}(0)\rangle $ in terms of the stationary state $|\psi_{HNstat,y} \rangle$ as
\begin{align}
|\psi_{2d}(0) \rangle =|\psi_{HNstat,y} \rangle+ \sqrt{\frac{8+l}{N}}\frac{1}{2} \biggl( |6  \rangle   + |7 \rangle  \biggr)\otimes |v_{x}; \sqrt{N}/2 \rangle\,.
\label{initialof2dHN}
\end{align}
The final state, after applying $U_{l}$ repeatedly $t$ times, becomes 
\begin{align}
U_{l}^{t}|\psi_{2d}(0) \rangle =|\psi_{HNstat,y} \rangle+ \sqrt{\frac{8+l}{N}}\frac{1}{2} U_{l}^{t}\Biggl[\biggl( |6 \rangle   + |7 \rangle  \biggr)\otimes |v_{x}; \sqrt{N}/2 \rangle \Biggr]
\label{time2dHN}
\end{align}

It is important to note that the first component on the right side of eq. \eqref{time2dHN} remains unchanged under the action of the evolution operator; only the second component is affected. The upper limit on the success probability of locating a marked vertex with a directed self-loop can be determined by setting $U_{l}^{t} = -\mathbb{I}$ in the second term. Consequently, the success probability for a marked vertex with a directed self-loop is constrained by $p_{2d}(t) \le \mathcal{O}(1/N)$. The same argument also holds  for  $U_{Grov}$. It suggests that the quantum walk search  conducted using $U_{l}$ and $U_{Grov}$ on a two-dimensional periodic lattice featuring additional long-range  edges are unable to identify a vertex with  directed self-loop, as the success probability constrained  to be limited by the initial success probability.  However, when  $U_{G}$ is applied to the initial state, both the first and second components of eq.  \eqref{initialof2dHN} undergo non-trivial transformations, since $|\psi_{HNstat,y} \rangle$ is no longer an eigenstate of $U_{G}$. Following multiple applications of the evolution operator, the final state achieves a significantly high and constant  overlap with the marked vertex.
In the case of $U_{SKW}$, both components of the initial state also transform non-trivially. However, to achieve a high success probability, it is essential to implement amplitude amplification in conjunction with the repeated application of $U_{SKW}$.

\subsubsection{Searching  for vertices  with  two directed self-loops of the long-range edges}

\begin{figure}[h!]
    \centering
    \includegraphics[scale=0.60]{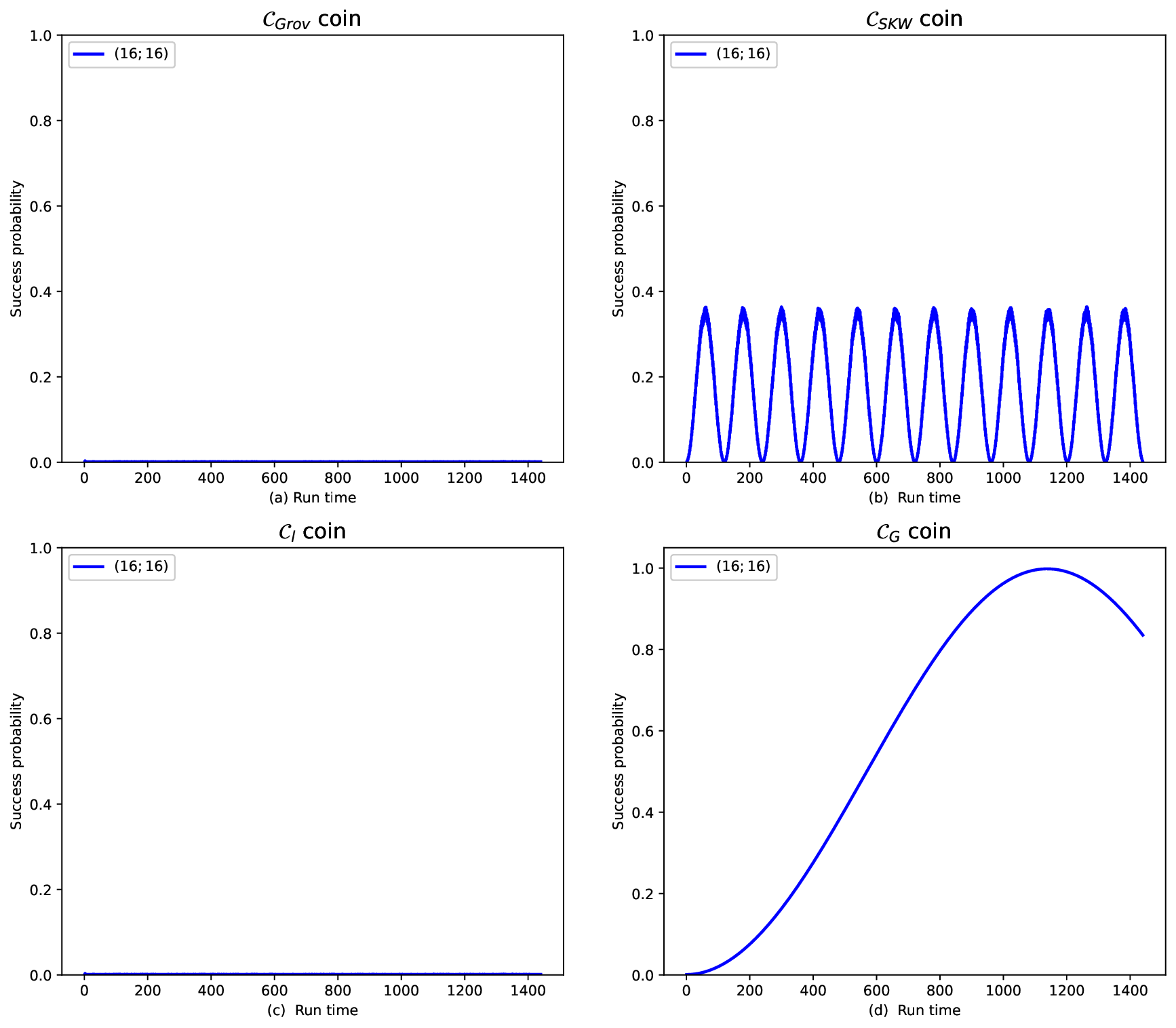}
    \caption{Success probability to measure a marked vertex $(16;16)$ on a  $2^5 \times 2^5$ two-dimensional periodic lattice with extra long-range  edges as a function of the number of iteration steps for (a) $\mathcal{C}_{Grov}$, (b) $\mathcal{C}_{SKW}$, (c) $\mathcal{C}_{l}$, and (d) $\mathcal{C}_{G}$  coins.}
  \label{2D-HN4}
\end{figure}

{\it Experimental results:} The behavior of the success probability to measure the  marked vertex at $(16;16)$, which has two directed self-loops  on a $2^5 \times 2^5$ lattice as a function of the number of iteration steps has been presented in FIG. \ref{2D-HN4}. We fix the self-loop weight at $l= 4.0/N$.   
We see that the success probability for the $\mathcal{C}_{Grov}$ and $\mathcal{C}_{l}$ coins  represented by FIGs. \ref{2D-HN4}(a) and \ref{2D-HN4}(c) respectively  do not grow at all even if we increase the number of iteration steps. For the $\mathcal{C}_{SKW}$ and $\mathcal{C}_{G}$ coins  success probability grow as a function of the number of iteration steps represented by figs. \ref{2D-HN4}(b) and \ref{2D-HN4}(d) respectively. For the SKW coin, as we can see, the success probability is not significantly high, so we need to  apply amplitude amplification to further enhance the success probability. 
Although the experimental result  only for $(16;16)$ is presented in FIG. \ref{2D-HN4},  experimental results for other exceptional  vertices $(16;32),(32;16)$ and $(32;32)$ are found to be similar.
The reason that these configurations cannot be searched by $\mathcal{C}_{Grov}$ and $\mathcal{C}_{l}$ coin is explained like previous subsection.

We consider the stationary state, which has large overlap with initial state of the quantum walk search. In this subsection, we construct the stationary state to the exceptional configuration occurred from two  directed self-loops vertex. 
The stationary state is the eigenvector of time evolution operator with unit eigenvalue. 
For the description of action of coin operators to the stationary state, we need the following pair of non-normalized coin states $|\psi_{c}^{\mathrm{le},x+} \rangle,|\psi_{c}^{\mathrm{le},x-} \rangle$, and $|\psi_{c}^{\mathrm{le},y+} \rangle, |\psi_{c}^{\mathrm{le},y-} \rangle$. These states are defined at previous subsections.
We consider the stationary state associated with exceptional vertices  $|\sqrt{N}/2 ;\sqrt{N}/2\rangle, |\sqrt{N}/2 ;\sqrt{N}\rangle, |\sqrt{N}; \sqrt{N}/2 \rangle$ and $|\sqrt{N}; \sqrt{N} \rangle$. The stationary states associated with $|\sqrt{N}/2; \sqrt{N}/2 \rangle$ is defined as follow
\begin{align}
|\psi_{HNstat,two} \rangle :=|\psi_{2d}(0) \rangle - \sqrt{\frac{8+l}{N}}\frac{1}{4} \biggl( |4  \rangle +|5  \rangle +|6  \rangle +|7  \rangle\biggr)\otimes | \sqrt{N}/2;\sqrt{N}/2 \rangle ,
\end{align}
where the initial state $|\psi_{2d}(0)\rangle$  is given by eq. (\ref{in2dl}).  The stationary states associated with $|\sqrt{N}/2;  \sqrt{N}\rangle, |\sqrt{N}; \sqrt{N}/2 \rangle$ and $| \sqrt{N}; \sqrt{N} \rangle$ are defined similarly.
The following  is the action of four evolution  operators on these stationary states:

\begin{enumerate}
\item $U_{l}$: We can demonstrate that the stationary states $|\psi_{HNstat,two} \rangle$ is eigenstate of the time evolution operator  $U_{l}=S\mathcal{C}_{l}$, with a unit eigenvalue
\begin{align}
    U_{l} |\psi_{HNstat,two} \rangle=|\psi_{HNstat,two} \rangle.
\end{align}

\item $U_{Grov}$: $|\psi_{HNstat,two} \rangle$ remains eigenstate for the evolution operator  $U_{Grov}= S\mathcal{C}_{Grov}=U_{l=0}$ with Grover coin 
\begin{align}
    U_{Grov} |\psi_{HNstat,two} \rangle&=|\psi_{HNstat,two} \rangle.
\end{align}
Here we take  $l=0$ limit.

\item $U_{G}$: However, under the operator $U_{G}=S\mathcal{C}_{G}$, the stationary states do not remain stationary, as illustrated by the following expression
\begin{eqnarray}\nonumber
    U_{G} |\psi_{HNstat,two}  \rangle = \hspace{13cm}\\ 
    |\psi_{HNstat,two}  \rangle -\frac{1}{2\sqrt{N}}S\Biggl[\biggl(|\psi_{c}^{\mathrm{le},x+} \rangle  +|\psi_{c}^{\mathrm{le},x-}+|\psi_{c}^{\mathrm{le},y+} \rangle  +|\psi_{c}^{\mathrm{le},y-} \rangle  \biggr) \otimes |\sqrt{N}/2; \sqrt{N}/2 \rangle \Biggl] \nonumber\\ 
    - 2\sqrt{\frac{l}{N(8+l)}}SD_{l}\biggl( |\mbox{lp} \rangle \otimes|\sqrt{N}/2; \sqrt{N}/2 \rangle   \biggr)\,. \hspace{6.5cm}
\end{eqnarray}

\item $U_{SKW}$: Similarly, under $U_{SKW}=S \mathcal{C}_{SKW}$, the stationary states also do not remain stationary, as shown by the following expression
\begin{align}
    U_{SKW} |\psi_{HNstat,two} \rangle   =|\psi_{HNstat,two} \rangle  -\frac{1}{2\sqrt{N}}S\Biggl[\biggl(|\psi_{c}^{\mathrm{le},x+} \rangle +|\psi_{c}^{\mathrm{le},x-} \rangle +|\psi_{c}^{\mathrm{le},y+} \rangle +|\psi_{c}^{\mathrm{le},y-} \rangle \biggr)\otimes |\sqrt{N}/2; \sqrt{N}/2 \rangle\Biggl].
\end{align}
\end{enumerate}
Note that, for the quantum walk search with SKW coin there is no self-loop, so we  set $l =0$ wherever necessary. 
We can express the initial state $|\psi_{2d}(0)\rangle $ in terms of the stationary state $|\psi_{HNstat,two} \rangle$ as
\begin{align}
|\psi_{2d}(0) \rangle =|\psi_{HNstat,two} \rangle+ \sqrt{\frac{8+l}{N}}\frac{1}{4} \biggl( |4  \rangle   + |5 \rangle+|6  \rangle   + |7 \rangle  \biggr)\otimes |\sqrt{N}/2; \sqrt{N}/2 \rangle\,.
\label{initialof2dtwo}
\end{align}
The final state, after applying $U_{l}$ repeatedly $t$ times, becomes 
\begin{align}
U_{l}^{t}|\psi_{2d}(0) \rangle =|\psi_{HNstat,two} \rangle+ \sqrt{\frac{8+l}{N}}\frac{1}{4} U_{l}^{t}\Biggl[\biggl(  |4  \rangle   + |5 \rangle+|6 \rangle   + |7 \rangle  \biggr)\otimes |\sqrt{N}/2; \sqrt{N}/2 \rangle \Biggr]
\label{time2dtwo}
\end{align}

It is important to note that the first component on the right side of eq. \eqref{time2dtwo} remains unchanged under the action of the evolution operator; only the second component is affected. The upper limit on the success probability of locating a marked vertex with a directed self-loop can be determined by setting $U_{l}^{t} = -\mathbb{I}$ in the second term. Consequently, the success probability for a marked vertex with a directed self-loop is constrained by $p_{2d}(t) \le \mathcal{O}(1/N)$. The same argument also holds  for  $U_{Grov}$. It suggests that the quantum walk search  conducted using $U_{l}$ and $U_{Grov}$ on a two-dimensional periodic lattice featuring additional long-range  edges are unable to identify a vertex with  directed self-loop, as the success probability constrained  to be limited by the initial success probability.  However, when  $U_{G}$ is applied to the initial state, both the first and second components of eq.  \eqref{initialof2dtwo} undergo non-trivial transformations, since $|\psi_{HNstat,y} \rangle$ is no longer an eigenstate of $U_{G}$. Following multiple applications of the evolution operator, the final state achieves a significantly high and constant  overlap with the marked vertex.
In the case of $U_{SKW}$, both components of the initial state also transform non-trivially. However, to achieve a high success probability, it is essential to implement amplitude amplification in conjunction with the repeated application of $U_{SKW}$.

\section{Conclusions} \label{con}

Quantum walk  has proven to be highly effective for searching for a single marked vertex on various graphs. However, when it comes to identifying multiple marked vertices, certain limitations arise. In particular, some widely used  coin operators struggle to locate specific exceptional configurations.
In this work, we explored one- and two-dimensional lattices augmented with extra long-range edges provided by the HN4 . In the case of a two-dimensional periodic lattice with these additional edges, the lackadaisical quantum walk with $\mathcal{C}_{G}$ coin achieves an optimal time complexity of $\mathcal{O}(\sqrt{N/M})$ for quantum search. Within these lattices, we identified new exceptional configurations and corresponding stationary states, which help explain why the coin operators $C_{Grov}$ and $C_{l}$ fail to locate these configurations.

In Section \ref{1D}, we investigated a one-dimensional periodic lattice with extra long-range edges from the HN4 . Here, we discovered exceptional configurations—specifically, a vertex with directed self-loops associated with the long-range edge. Numerical evidence indicates that this configuration is indeed exceptional, as it cannot be reliably detected by $C_{Grov}$ and $C_{l}$. In contrast, $C_{SKW}$ and $C_G$ are able to search for this configuration, with $C_G$ achieving high success probabilities on the order of $\mathcal{O}(1)$. We further constructed a stationary state linked to this new exceptional configuration, providing a theoretical explanation for the observed failure of $C_{Grov}$ and $C_{l}$.

Section \ref{2D} is devoted to the study of a two-dimensional periodic lattice with extra long-range edges from the HN4. Our numerical investigations show that configurations previously deemed exceptional on a regular two-dimensional periodic lattice (such as diagonal configurations) are no longer exceptional when long-range edges are introduced—owing to the absence of associated stationary states. Instead, new exceptional configurations emerge: a pair of vertices that are adjacent via long-range edges, a vertex with one directed self-loop of a long-range edge, and a vertex with two directed self-loops of the long-range edges. Our results indicate that these configurations cannot be found by $C_{Grov}$ and $C_{l}$, yet they can be efficiently searched using $C_{SKW}$ and $C_G$, with $C_G$ once again achieving high success probabilities. We also constructed stationary states corresponding to these new exceptional configurations.

Looking ahead, it would be interesting to explore  potential exceptional configurations in two-dimensional lattices with extra long-range edges provided by the combination of  HN3 and HN4. Another avenue for further research is to derive analytical expressions for the success probabilities and the runtime performance of $C_{SKW}$ and $C_G$. Achieving this will likely require finding the eigenvectors of the time evolution operators—in addition to the stationary states (with eigenvalue one)—associated with these coin operators. Such endeavors promise to deepen our understanding of these lattices. Additionally, investigating applications of these lattices in other areas of information science-as well as exploring the potential of $C_G$ in other problems-represents a promising direction for future research.

\vspace{1cm}

Data availability Statement:  The data and database information  generated during and/or analyzed during the current study are included in the article.

\vspace{0.5cm}

Conflict of interest: The authors have no competing interests to declare that are relevant to the content of this article. 


\end{document}